\documentclass[12pt]{article}
\usepackage{fullpage}
\usepackage{float}
\usepackage{afterpage}
\usepackage{amsmath}
\usepackage{graphicx,psfrag,epsf}
\usepackage{enumerate}
\usepackage{caption}
\usepackage{natbib}
\usepackage{amssymb}
\usepackage{graphicx}
\usepackage{amsfonts}
\usepackage{amsmath}
\usepackage{amsthm}
\usepackage{nameref}

\DeclareRobustCommand{\stir}{\genfrac\{\}{0pt}{}}
\usepackage[ruled,vlined,boxed]{algorithm2e}

\newcommand{\blind}{0}

\begin{document}

\makeatletter
\setlength{\@fptop}{0pt}
\setlength{\@fpbot}{0pt plus 1fil}
\makeatother

\def\spacingset#1{\renewcommand{\baselinestretch}%
{#1}\small\normalsize} \spacingset{1}

\if0\blind
{
  \title{\bf Constrained Community Detection\\in Social Networks}
  \author{Weston D. Viles\\
    Department of Biomedical Data Science\\
    Geisel School of Medicine\\Dartmouth College\\
    Hanover, NH 03755\\
    \texttt{weston.d.viles@dartmouth.edu}\\
    and\\
    A. James O'Malley\\Department of Biomedical Data Science\\
    The Dartmouth Institute\\Geisel School of Medicine\\Dartmouth College\\
    Hanover, NH 03755\\\texttt{james.omalley@dartmouth.edu}}
    \date{}
  \maketitle
} \fi

\if1\blind
{
  \bigskip
  \bigskip
  \bigskip
  \begin{center}
    {\LARGE\bf Constrained Community Detection\\in Social Networks}
\end{center}
  \medskip
} \fi

\pagenumbering{gobble}

\newpage

\begin{center}
\textbf{Author's Footnote}

\bigskip

\if0\blind
{
\begin{minipage}{.9\textwidth}
\spacingset{1.3}
Weston D. Viles is a Postdoctoral Research Associate, Department of Biomedical Data Science, Geisel School of Medicine at Dartmouth College, HB 7261, One Medical Center Drive, Lebanon, NH 03756 (e-mail: weston.d.viles@dartmouth.edu) and A. James O'Malley is Professor of Biostatistics in the Department of Biomedical Data Science and in the Dartmouth Institute for Health Policy and Clinical Practice, Geisel School of Medicine at Dartmouth College, HB 7251, One Medical Center Drive Lebanon, NH 03756 (e-mail: james.omalley@dartmouth.edu).  W.D.V. is partially supported by the National Institutes of Health training grant R25CA134286 and A.J.O. by NIH/NIA grant U01AG046830.
\end{minipage}
} \fi

\if1\blind
{
\phantom{\begin{minipage}{.9\textwidth}
\spacingset{1.3}
Weston D. Viles is a Postdoctoral Research Associate, Department of Biomedical Data Science, Geisel School of Medicine at Dartmouth College, HB 7261, One Medical Center Drive, Lebanon, NH 03756 (e-mail: weston.d.viles@dartmouth.edu) and A. James O'Malley is Professor of Biostatistics in the Department of Biomedical Data Science and in the Dartmouth Institute for Health Policy and Clinical Practice, Geisel School of Medicine at Dartmouth College, HB 7251, One Medical Center Drive Lebanon, NH 03756 (e-mail: james.omalley@dartmouth.edu).  W.D.V. is partially supported by the National Institutes of Health training grant R25CA134286 and A.J.O. by NIH/NIA grant U01AG046830.
\end{minipage}}
} \fi
\end{center}

\vspace{.025\paperheight}

\begin{center}
\textbf{Abstract}

\bigskip
\begin{minipage}{.9\textwidth}
\spacingset{1.3}
Community detection in networks is the process of identifying unusually well-connected sub-networks and is a central component of many applied network analyses.  The paradigm of modularity optimization stipulates a partition of the network's vertices which maximizes the difference between the fraction of edges within groups (communities) and the expected fraction if edges were randomly distributed.  The modularity objective function incorporates the network's topology exclusively and has been extensively studied whereas the integration of constraints or external information on community composition has largely remained unexplored.  We impose a penalty function on the modularity objective function to regulate the constitution of communities and apply our methodology in identifying health care communities (HCCs) within a network of hospitals such that the number of cardiac defibrillator surgeries performed within each HCC exceeds a minimum threshold.  This restriction permits meaningful comparisons in cardiac care among the resulting health care communities by standardizing the distribution of cardiac care across the hospital network.
\end{minipage}
\end{center}

\noindent%
\vfill

\newpage

\pagenumbering{arabic}

\spacingset{1.45} 
\setcounter{section}{1}
\section*{\centering \arabic{section}. \uppercase{Introduction}}

Networks are collections of interconnected entities that are frequently represented by graphs encoding the pairwise relationships within the interacting systems.  Network science and graph theory provide methodology for analyzing relational data from a variety of scientific disciplines.  As a pursuit of network science, community detection is the process of identifying exceptionally dense subnetworks of mutually well-connected entities, known as communities, that often have functional meaning in the network.  Notable approaches to community detection include the clique percolation method \citep{Palla2005}, spectral partitioning \citep{Chung1997}, degree-corrected stochastic block models \citep{Karrer2011}, modularity optimization \citep{Newman2006}, and multi-slice network community detection \citep{Mucha2010}.  These approaches are designed for the unsupervised partitioning of the vertex set of a graph into unusually cohesive subsets of vertices, and with varied applications in sociology \citep{Mucha2007}, computer architecture \citep{Kolda2000}, and biology \citep{Andrade2011}, illustrate that community detection procedures are applied broadly in scientific research.

Community detection procedures commonly integrate network connectivity without regard for other quantities of interest, e.g., auxiliary measurements on vertices.  Our motivating problem is to partition a nation-wide network of hospitals into subnetworks that (i) exhibit a high level of within-group patient sharing and (ii) collectively have hosted a minimum number of \textit{implantable cardioverter defibrillator} (ICD) surgeries.  We utilize data on health insurance claims made to the Medicare national social insurance program during the period 2006-2011 in addition to the quantity of ICD surgeries at major cardiovascular referral centers, formerly known as \textit{cardiac care facilities} (CCFs).  The preeminent existing work in this domain is the \textit{Dartmouth Atlas} in which \cite{Wennberg1995} assigned hospitals to one of 306 \textit{health referral regions} (HRRs) representing markets for tertiary medical care.  The significant contributions made by the Dartmouth Atlas to health services research motivates our work but our methodology, which is based on network-graph topology, is a departure from the HRRs adherence to geographic proximity.

We represent the nation-wide hospital network with the graph $G=(\mathbf{V},\mathbf{E}_{\mathbf{W}})$ which designates one vertex $v\in\mathbf{V}$ for each hospital and a non-negatively weighted edge $\{v,u,W_{uv}\}\in\mathbf{E}_\mathbf{W}$ for each pair $\{u,v\}\in\mathbf{V}^2$ of vertices, where $\mathbf{V}^2=\mathbf{V}\times\mathbf{V}$ is the set of vertex pairs.  The weight $W_{uv}$ of edge $\{v,u,W_{uv}\}$ reflects the quantity of shared patient visits recorded in Medicare claims data between the hospitals associated with vertices $u,v\in\mathbf{V}$. Additionally, we denote the subset of \textit{special vertices} $\mathbf{V}^\prime\subseteq\mathbf{V}$ to correspond with the subset of hospitals that are CCFs, i.e., hospitals at which at least one ICD surgery was performed. The number of ICD surgeries performed at a hospital is information not encoded in the graph $G$ and, therefore, its integration into a standard community detection procedure must be made explicitly.

We define a \textit{health care community assignment} as the designation of hospitals to communities which optimizes network \textit{modularity}, the difference between the fraction of (weighted) edges within communities and the expected fraction if edges were randomly distributed, subject to the constraint that the total volume of ICD surgeries performed at hospitals within each community exceeds a specified minimum.  This setup defines a constrained optimization problem. Specifically, we seek to maximize modularity while requiring that the number of ICD surgeries per \textit{health care community} (HCC) exceeds the specified minimum volume of ICD surgeries $\tau\geq0$: a restriction we will encode with a binary variable indicating the feasibility of a community assignment.  

Existing work in this realm considers incorporating additional information in the form of individual entity labels and pairwise constraints, i.e. that two vertices must be labeled similarly or differently (a different constraint from our own) in modularity optimization, see \cite{Eaton2012}.  A restriction of the type we consider here has, to our knowledge, remained unstudied.  In the context of the hospital network, the HRRs defined previously associated local health care markets to the tertiary care facilities where the plurality of the residents were referred for major cardiac procedures.  An HRR is a reflection of its regional health care market and, because necessarily within each is a hospital specialized in cardiac surgery, a comparison across regions is facilitated.  In an effort to standardize cardiac care among health care subnetworks, we present an initial undertaking towards developing a general paradigm of constrained community detection.  In particular, because constraint satisfaction in general optimization problems provides context, the health care communities identified by our constrained community detection approach have real-world utility.

The organization of this article is as follows.  In Section \ref{funcoptim}, we mathematically formulate the constrained optimization problem.  In Section \ref{penrep}, we transform the  constrained optimization problem into a penalized optimization problem with remarks about computational complexity in the appendix.  In Section \ref{optproc}, we present our algorithmic procedure for estimating the solution to the penalized optimization problem. Utilizing our procedure in Section \ref{appl}, we estimate the health care communities of the nation-wide hospital network.  We conclude with possible extensions and future research in Section \ref{extrem}.

\addtocounter{section}{1}
\section*{\centering \arabic{section}. \uppercase{Problem Specification}}
\addtocounter{section}{-1}
\refstepcounter{section}
\label{funcoptim}

The graph $G=(\mathbf{V},\mathbf{E}_W)$ encodes patterns of patient-sharing among the $p=4734$ hospitals participating in the Medicare social insurance system.  The quantity of patient-sharing among hospitals is represented in the (symmetric) weighted adjacency matrix $\mathbf{W}=[W_{uv}]$, for $u,v\in\mathbf{V}$, in which $W_{uv}>0$ if a patient sought care at both of the hospitals corresponding to vertices $u,v\in\mathbf{V}$ and, otherwise, $W_{uv}=0$.  The modularity function $Q(\mathbf{x}|G)$ assesses the extent to which a partition of network vertices corresponds to the densely-connected vertex subsets in the network and is defined as 
\begin{equation}\label{mod}
Q(\mathbf{x}|G) = \frac{1}{2m}\sum_{\{u,v\}\in\mathbf{V}^2}\left(W_{uv}-\frac{d_ud_v}{2m}\right)1\{x_u=x_v\},
\end{equation}
where $d_u=\sum_vW_{uv}$ is the \textit{degree} of vertex $u$, $2m=\sum_ud_u$ is the total weight of the network ($m$ is the number of edges in the network if $G$ is unweighted), and $x_u$ is the community assignment for vertex $u\in\mathbf{V}$. The vector $\mathbf{x}\in\mathbf{K}_p$ is the network-wide \textit{vertex community assignment} and is an element of the set $\mathbf{K}_p=\left\{0,1,\ldots,p-1\right\}^p$ that characterizes all possible community designations for vertices in the network.  The \textit{maximum} modularity $Q(\mathbf{x}^{mod}|G)$ of the hospital network is
\begin{equation}\label{modopt}
Q(\mathbf{x}^{mod}|G) = \max_{\mathbf{x}\in\mathbf{K}_p}\hspace{1mm}Q(\mathbf{x}|G),
\end{equation}
where $\mathbf{x}^{mod}$ is a vertex assignment achieving modularity optimality.  Existing computational procedures for estimating $\mathbf{x}^{mod}$, e.g. simulated annealing \citep{Metropolis1953} and the Louvain method \citep{Blondel2008}, have been developed in lieu of intractable enumeration of $O([p/\log(p)]^p)$ vertex assignments (see the appendix for a brief complexity explanation).  These unconstrained optimization methods yield high-modularity vertex assignments on the nation-wide hospital network, but since these estimates of $\mathbf{x}^{mod}$ do not satisfy the conditions for health care community assignments for any $\tau\geq0$ in the hospital network, we enforce satisfying the constraint that a sufficient volume of ICD surgeries were performed by the hospitals constituting each health care community with our explicitly constrained procedure in Section \ref{appl}.

We associate with each vertex $v\in\mathbf{V}$ the function $f:\mathbf{V}\mapsto\mathbb{Z}^+\cup\{0\}$, where $f(v)$ is the number of ICD surgeries performed at the hospital associated with $v\in\mathbf{V}$, and specify that $f(v)>0$ for $v\in\mathbf{V}^\prime$ whereas $f(v)=0$ for $v\in\mathbf{V}\setminus\mathbf{V}^\prime$.  We define
\begin{equation}\nonumber
	F(v,\mathbf{x}) = \sum_{u\in\mathbf{V}}f(u)1\{x_v=x_u\}
\end{equation}
as the number of ICD surgeries performed in the community to which vertex $v$ belongs.  It is straightforward that for $u,v\in\mathbf{V}$, if $x_v=x_u$ then $F(u,\mathbf{x})=F(v,\mathbf{x})$.  Note that if the community assignment $\mathbf{x}$ satisfies the constraint, then $F(v,\mathbf{x})>\tau$ for all $v\in\mathbf{V}$.  We define the binary variable $T_\tau(\mathbf{x},\mathbf{V})=1\{\min_{v\in\mathbf{V}}F(v,\mathbf{x})\leq\tau\}$ to indicate the event that the community assignment $\mathbf{x}$ is in violation of the community-wide total ICD surgeries minimum volume constraint.  That is, $T_\tau(\mathbf{x},\mathbf{V})=1$ if there exists a vertex $v\in\mathbf{V}$ assigned to community $x_v$ for which the total number of ICD surgeries $F(v,\mathbf{x})$ performed in hospitals assigned to $x_v$ does not exceed $\tau$ and, otherwise, $T_\tau(\mathbf{x},\mathbf{V})=0$.  For some $\tau\geq0$, we define
\begin{equation}\nonumber
\mathbf{S}_{\tau}(\mathbf{V}) = \left\{\mathbf{x}\in\mathbf{K}_p:T_\tau(\mathbf{x},\mathbf{V})=0\right\}
\end{equation}
as the feasible subset $\mathbf{S}_{\tau}(\mathbf{V})\subseteq\mathbf{K}_p$ of vertex assignments.  In the special case of $\tau=0$, the variable $T_0(\mathbf{x},\mathbf{V})$ indicates the event that there exists a community in the network to which no special vertex $v\in\mathbf{V}^\prime$ belongs, i.e., no hospital performing cardiac surgery (CCF) belongs.

Our objective is to identify the maximum modularity solution restricted to the space of feasible vertex assignments $\mathbf{S}_{\tau}(\mathbf{V})$.  The {maximum feasible} modularity $Q(\mathbf{x}^*|G)$ of the hospital network is defined as
\begin{equation}\label{optim}
Q(\mathbf{x}^*|G) = \max_{\mathbf{x}\in\mathbf{S}_{\tau}(\mathbf{V})}Q\left(\mathbf{x}|G\right),
\end{equation}
where $\mathbf{x}^*$ is the vertex assignment achieving constrained modularity optimality.  We address estimating a solution of the formal constrained modularity optimization problem (CMOP)
\begin{equation}\label{CMOP}\nonumber
	\mbox{CMOP: maximize $Q(\mathbf{x}|G)$ over $\mathbf{x}\in\mathbf{K}_p$ subject to $T_\tau(\mathbf{x},\mathbf{V})=0$}
\end{equation}
\noindent with our forthcoming constrained optimization procedure.

\addtocounter{section}{1}
\section*{\centering \arabic{section}. \uppercase{Transformation to Penalized Modularity}}
\addtocounter{section}{-1}
\refstepcounter{section}
\label{penrep}

We transform the CMOP to an equivalent penalized optimization problem more amenable to computational evaluation.  The maximum feasible modularity defined in Equation \eqref{optim} is equivalently
\begin{equation}\label{optim2}
	Q(\mathbf{x}^*|G)=\max_{\mathbf{x}\in\mathbf{K}_p}\left\{Q(\mathbf{x}|G)-Q(\mathbf{x}^{mod}|G)T_\tau(\mathbf{x},\mathbf{V})\right\}
\end{equation}
since any $\mathbf{x}\notin\mathbf{S}_\tau(\mathbf{V})$ will incur a penalty guaranteed to demote its optimality.  Note that the constraint function $T_\tau(\mathbf{x}|\mathbf{V})$ can be expressed as
\begin{equation}\label{defT}\nonumber
	T_\tau(\mathbf{x}|\mathbf{V}) = \min\left\{1,\sum_{v\in\mathbf{V}}1\{F(v,\mathbf{x})\leq\tau\}\right\},
\end{equation}
where the sum counts the number of vertices assigned to communities in violation of the constraint.  Equation \eqref{optim2} is thus amenable to vertex-level indicators of infeasibility and is equivalent to
\begin{equation}\label{optim4}
	Q(\mathbf{x}^*|G) = \max_{\mathbf{x}\in\mathbf{K}_p}\left\{Q(\mathbf{x}|G)-\lambda\sum_{v\in\mathbf{V}}\lambda_v1\{F(v,\mathbf{x})\leq\tau\}\right\},
\end{equation}
where $\lambda>0$ and $\lambda_v>0$, for $v\in\mathbf{V}$, are great enough to guarantee that the maximum $\mathbf{x}^*\in\mathbf{S}_\tau(\mathbf{V})$.  The contribution of terms associated with vertex $v\in\mathbf{V}$ toward the modularity objective function $Q(\mathbf{x}|G)$ as defined in Equation \eqref{mod}, is
\begin{equation}\nonumber
Q(x_v|\mathbf{x}_{-v},G) = \frac{1}{2m}\left[\left(W_{vv}-\frac{d_v^2}{2m}\right)+2\sum_{\substack{u\in\mathbf{V}\\ u\neq v}}\left(W_{uv}-\frac{d_ud_v}{2m}\right)1\{x_u=x_v\}\right],
\end{equation}
where $\mathbf{x}_{-v}$ are vertex community labels not associated with vertex $v$.  The total interaction weight associated with vertex $v$ is
\begin{equation}\nonumber
\frac{1}{m}\sum_{\substack{u\in\mathbf{V}\\ u\neq v}}\left(W_{uv}-\frac{d_ud_v}{2m}\right) = -\frac{1}{m}\left(W_{vv}-\frac{d_v^2}{2m}\right)
\end{equation}
and, thus, we weight the contribution of the penalty term in Equation \eqref{optim4} associated with vertex $v$ by $\lambda_v=\frac{1}{m}\left|W_{vv}-\frac{d_v^2}{2m}\right|$ to balance its associated total interaction weight.  The object optimized in Equation \eqref{optim4} is thus
\begin{eqnarray}
\mathcal{H}_{\tau,\lambda}(\mathbf{x}|G) &=& \nonumber \frac{1}{m}\left[\frac{1}{2}\sum_{\{u,v\}\in\mathbf{V}^2}\left(W_{uv}-\frac{d_ud_v}{2m}\right)1\{x_u=x_v\}\right.\\
& &\label{hamil} \quad\left.-\lambda\sum_{v\in\mathbf{V}}\left|W_{vv}-\frac{d_v^2}{2m}\right|1\{F(v,\mathbf{x})\leq\tau\}\right].
\end{eqnarray}
We have chosen to express the function $\mathcal{H}_{\tau,\lambda}(\mathbf{x}|G)$ in the form given in Equation \eqref{hamil} to reflect its correspondence to the \textit{Hamiltonian} of a Potts model with \textit{interaction component}
\begin{equation}\nonumber
\mathcal{H}^I(\mathbf{x}|G) = \frac{1}{2m}\sum_{\{u,v\}\in\mathbf{V}^2}\left(W_{uv}-\frac{d_ud_v}{2m}\right)1\{x_u=x_v\}
\end{equation}
and \textit{external field}
\begin{equation}\nonumber
\mathcal{H}^E_{\tau,\lambda}(\mathbf{x}|\mathbf{V}) = -\frac{\lambda}{m}\sum_{v\in\mathbf{V}}\left|W_{vv}-\frac{d_v^2}{2m}\right|1\{F(v,\mathbf{x})\leq\tau\},
\end{equation}
such that $\mathcal{H}_{\tau,\lambda}(\mathbf{x}|G)=\mathcal{H}^I(\mathbf{x}|G)+\mathcal{H}^E_{\tau,\lambda}(\mathbf{x}|\mathbf{V})$.  The maximum feasible modularity vertex community assignment is thus defined, in a manner equivalent to Equation \eqref{optim}, as
\begin{equation}\label{optham}
\mathbf{x}^* = \underset{\mathbf{x}\in\mathbf{K}_p}{\arg\max}\hspace{1mm}\mathcal{H}_{\tau,\lambda}(\mathbf{x}|G),
\end{equation}
for which $\lambda\geq1$ guarantees $\mathbf{x}^*\in\mathbf{S}_\tau(\mathbf{V})$, in accordance with Equation \eqref{optim2}.  Therefore, $\mathbf{x}^*$ of Equation \eqref{optham} is the solution to the CMOP.

\addtocounter{section}{1}
\section*{\centering \arabic{section}. \uppercase{Constrained Optimization Procedure}}
\addtocounter{section}{-1}
\refstepcounter{section}
\label{optproc}

Modularity optimization is a combinatorial problem in the intractable class NP \citep{Brandes2008}.  Common approaches for identifying high-modularity community assignments that approximate $\mathbf{x}^{mod}$ in Equation \eqref{modopt} involve greedy or, more generally, stochastic optimization.  The deterministic Louvain method \citep{Blondel2008} efficiently approximates the modularity of large networks by alternating between two sub-routines, the first of which involves iteratively reassigning vertices to the communities which most increase the modularity of the overall vertex community assignment.  When no vertex label can be reassigned to increase modularity, the second sub-routine collapses communities into individual vertices (in a newly constructed \textit{folded graph}) that are connected by edge weights equal to the total edge weight between the communities, see Section \ref{foldproc}.  Subsequent greedy optimization of vertex community assignments on the folded graph are carried out as in the previous step on the original graph.  This sequence of greedy optimization and folding is alternated until convergence.  Alternatively, simulated annealing \citep{Metropolis1953} is a Monte Carlo procedure  adapted from Metropolis-Hastings \citep{Minh2015} which generalizes greedy optimization by permitting sub-optimal vertex community label reassignments selected stochastically.  Candidate vertex community assignments are randomly chosen and accepted or rejected with probability determined by the proposed increase in modularity and the duration of execution of the procedure.  We propose a Monte Carlo procedure for optimizing $\mathcal{H}_{\tau,\lambda}(\mathbf{x}|G)$ in Equation \eqref{hamil} that follows the (i) local optimization and (ii) folding paradigms from each of these procedures in the following sections.  We first define the Monte Carlo transition probabilities for our stochastic, constrained optimization method and subsequently describe their implementation.

\subsection{The Probability Distribution of Community Assignments}\label{model}

We utilize the Hamiltonian $\mathcal{H}_{\tau,\lambda}(\mathbf{x}|G)$ in Equation \eqref{hamil} to equip the probability distribution that defines the transition probabilities for our Markov Chain Monte Carlo (MCMC) procedure.  The probability distribution is supported on the space $\mathbf{K}_p$ of vertex community assignments according to
\begin{eqnarray}
\mathbb{P}_{\tau,\lambda,\theta}(\mathbf{x}|G) &\propto&\nonumber \exp\left\{\theta\mathcal{H}_{\tau,\lambda}(\mathbf{x}|G)\right\}\\
&\propto&\nonumber \exp\left\{\theta\left[\mathcal{H}^I(\mathbf{x}|G)+\mathcal{H}^E_{\tau,\lambda}(\mathbf{x}|\mathbf{V})\right]\right\}\\
&\propto&\nonumber \exp\left\{\frac{\theta}{m}\left[\frac{1}{2}\sum_{\{u,v\}\in\mathbf{V}^2}\left(W_{uv}-\frac{d_ud_v}{2m}\right)1\{x_u=x_v\}\right.\right.\\
& &\label{premodel} \left.\left.
\phantom{\frac{1}{2}\sum_{\{u,v\}\in\mathbf{V}^2}}
-\lambda\sum_{v\in\mathbf{V}}\left|W_{uv}-\frac{d_v^2}{2m}\right|1\{F(v,\mathbf{x})\leq\tau\}\right]\right\},
\end{eqnarray}
where $\theta>0$ is a free parameter, known as the \textit{inverse temperature} of Boltzmann distributions (see Section \ref{locopt}), that interpolates $\mathbb{P}_{\tau,\lambda,\theta}(\mathbf{x}|G)$ between (i) the uniform distribution when $\theta\rightarrow0$ and (ii) a point mass when $\theta\rightarrow\infty$.  Since the optimal feasible community assignment $\mathbf{x}^*$ in Equation \eqref{optham} is the greatest mode of $\mathbb{P}_{\tau,\lambda,\theta}(\mathbf{x}|G)$, one may equivalently optimize Equation \eqref{premodel}.

When sampling from $\mathbb{P}_{\tau,\lambda,\theta}(\mathbf{x}|G)$, the interaction component $\mathcal{H}^I(x_v|\mathbf{x}_{-v},G)$ in the exponent of $\mathbb{P}_{\tau,\lambda,\theta}(x_v|\mathbf{x}_{-v},G)$ up-weights the probability of drawing candidate labels favoring modularity maximization while the external field $\mathcal{H}^E_{\tau,\lambda}(x_v|\mathbf{x}_{-v},\mathbf{V})$ down-weights the probability of drawing candidate labels in violation of the constraint.  In addition, we specify a \textit{weight function} $\mathcal{W}_{\tau,\gamma}(\mathbf{x}|G)$ on the community labels of special vertices $v\in\mathbf{V}^\prime$ that promotes their diffusion across communities in constraint-satisfying assignments.  Accordingly, we define the weight function
\begin{equation}\label{prior}
\mathcal{W}_{\tau,\gamma}(\mathbf{x}|G)\propto\exp\left\{-\gamma T_\tau(\mathbf{x}|\mathbf{V})\sum_{v\in\mathbf{V}^\prime}\gamma_v1\{F(v,\mathbf{x})-f(v)>\tau\}\right\},
\end{equation}
where $\gamma>0$ and $\gamma_v>0$ for $v\in\mathbf{V}^\prime$, and  $1\{F(v,\mathbf{x})-f(v)>\tau\}$ indicates the event that the quantity $f(v)$ of ICD surgeries associated with the CCF corresponding to special vertex $v\in\mathbf{V}^\prime$ contribute excess surgeries to the community labeled $x_v$.  In the event that some community in the network is in violation of the constraint, as indicated by $T_\tau(\mathbf{x}|\mathbf{V})$, the number of special vertices contributing excess ICD surgeries to their respective communities is totaled up to the scaling factors $\gamma_v$, by the sum in Equation \eqref{prior}.  Thus, for example, if $T_\tau(\mathbf{x}|\mathbf{V})=1$ and $1\{F(v,\mathbf{x})-f(v)>\tau\}=1$, for some $v\in\mathbf{V}^\prime$, then special vertex $v$ may be reassigned and contribute its $f(v)$ ICD surgeries elsewhere to a community in violation of the constraint without jeopardizing the feasibility of the community from where it came.

We set $\gamma=\lambda\theta/m$ and $\gamma_v=\lambda_v$ and write the full probability distribution of community labels $\mathbb{P}^*_{\tau,\lambda,\theta}(\mathbf{x}|G)$ as
\begin{eqnarray}
	\mathbb{P}^*_{\tau,\lambda,\theta}(\mathbf{x}|G) &\propto&\nonumber \mathbb{P}_{\tau,\lambda,\theta}(\mathbf{x}|G)\mathcal{W}_{\tau,\lambda\theta/m}(\mathbf{x}|G)\\
	&\propto&\nonumber \exp\{\theta\left[\mathcal{H}^I(\mathbf{x}|G)+\mathcal{H}^E_{\tau,\lambda}(\mathbf{x}|\mathbf{V})\right\}\mathcal{W}_{\tau,\lambda\theta/m}(\mathbf{x}|G)\\
	&\propto&\nonumber \exp\left\{\frac{\theta}{m}\left[\frac{1}{2}\sum_{\{u,v\}\in\mathbf{V}^2}\left(W_{uv}-\frac{d_ud_v}{2m}\right)1\{x_u=x_v\}\right.\right.\\
	& & \label{fullmodel} \left.\left.\phantom{\exp}-\lambda\sum_{v\in\mathbf{V}}\left|W_{uv}-\frac{d_v^2}{2m}\right|\chi(v,\mathbf{x})\right]\right\},
\end{eqnarray}
where the individual penalty terms (summands) of the external field $\mathcal{H}^E_{\tau,\lambda}(\mathbf{x}|G)$ and weight function $\mathcal{W}_{\tau,\lambda\theta/m}(\mathbf{x}|G)$ are combined into
\begin{equation}\nonumber
	\chi(v,\mathbf{x}) = 1\{F(v,\mathbf{x})>\tau\}+T_\tau(\mathbf{x}|\mathbf{V})1\{\{v\in\mathbf{V}^\prime\}\cap\{F(v,\mathbf{x})-f(v)>\tau\}\}.
\end{equation}
We present our vertex community assignment resampling procedure for generating samples near modes of $\mathbb{P}^*_{\tau,\lambda,\theta}(\mathbf{x}|G)$ in the following section.

\subsection{Local Optimization}\label{locopt}

Our MCMC optimization procedure is driven by the conditional distributions of $\mathbb{P}^*_{\tau,\lambda,\theta}(\mathbf{x}|G)$ in Equation \eqref{fullmodel}.  These distributions $\mathbb{P}^*_{\tau,\lambda,\theta}(x_v|\mathbf{x}_{-v},G)$ of label $x_v$ on the remaining labels $\mathbf{x}_{-v}$ take the form
\begin{eqnarray}
\mathbb{P}^*_{\tau,\lambda,\theta}(x_v|\mathbf{x}_{-v},G) &\propto&\nonumber \exp\left\{\theta\left[\mathcal{H}^I(x_v|\mathbf{x}_{-v},G)+\mathcal{H}^E_{\tau,\lambda}(x_v|\mathbf{x}_{-v},\mathbf{V})\right]\right\}\mathcal{W}_{\tau,\lambda\theta/m}(x_v|\mathbf{x}_{-v}|G)\\ &\propto&\nonumber \exp\left\{\frac{\theta}{m}\left[\sum_{\substack{u\in\mathbf{V}\\ u\neq v}}\left(W_{uv}-\frac{d_ud_v}{2m}\right)1\{x_u=x_v\}\right.\right.\\
& &\label{multinom} \left.\left.\phantom{\exp}-\lambda\left|W_{vv}-\frac{d_v^2}{2m}\right|\chi(v,\mathbf{x})\vphantom{\sum_{\substack{u\in\mathbf{V}\\ u\neq v}}}\right]\right\}.
\end{eqnarray}
The conditional distributions in Equation \eqref{multinom} define multinomial probabilities from which candidate vertex labels are iteratively drawn throughout  the iterative procedure described in Algorithm \ref{alg1}.
\afterpage{\IncMargin{1em}
\begin{algorithm}[htpb]
\spacingset{1}
\SetKwData{Left}{left}\SetKwData{This}{this}\SetKwData{Up}{up}
\SetKwFunction{Union}{Union}\SetKwFunction{FindCompress}{FindCompress}
\SetKwInOut{Input}{input}\SetKwInOut{Output}{output}
\Input{$G=(\mathbf{V},\mathbf{E}_W)$: Weighted network-graph\\
$\lambda$: Penalty weight parameter; $\theta$: Inverse temperature parameter\\
$\tau$: ICD volume threshold\\
$T$: Number of iterations}
\Output{$\mathbf{X}$: matrix of community labels}
\BlankLine
$\mathbf{x}\leftarrow[0,1,\ldots,|\mathbf{V}|-1]$\tcp*{array of $0,1,\ldots,|\mathbf{V}|-1$}
$\mathbf{X}\leftarrow\mbox{matrix}(0,|\mathbf{V}|,T)$\tcp*{$|\mathbf{V}|\times T$ matrix of zeros}
$\mathbf{X}[,1]\leftarrow\mathbf{x}$\;
\For{$t\leftarrow 2$ \KwTo $T$}{
$\theta\leftarrow\mbox{CoolingSchedule}(\theta,t)$\tcp*{default: constant}
$vertSeq\leftarrow\mbox{RandomPermute}(1:|\mathbf{V}|)$\tcp*{sequence of vertices}
\For{$k\in vertSeq$}{
$\mathbf{U}\leftarrow\mbox{unique}(\mathbf{x})$\tcp*{unique community labels}
$pr\leftarrow\mbox{array}(0,|\mathbf{U}|)$\tcp*{array of $|\mathbf{U}|$ zeros}
\For{$j\leftarrow1$ \KwTo $|\mathbf{U}|$}{
	$pr[j]\leftarrow\mathbb{P}^*_{\theta,\tau,\lambda}(\mathbf{U}[j],\mathbf{x}[-k],G)$\tcp*{compute conditional prob.}
}
$\mathbf{x}[k]\leftarrow\mathbf{U}[\mbox{RandomMultinomial}(pr)]$\tcp*{(i) draw from multinomial}
\DontPrintSemicolon\tcp*{(ii) assign label to $\mathbf{x}[k]$}
}
$\mathbf{X}[,t]\leftarrow\mathbf{x}$\tcp*{store labels in $\mathbf{X}$}
}
\Return $\mathbf{X}$\;
\caption{Local Optimization}
\label{alg1}
\end{algorithm}\DecMargin{1em}}
The inputs to Algorithm \ref{alg1} are the network-graph $G$ and the parameters of $\mathbb{P}^*_{\tau,\lambda,\theta}(\mathbf{x}|G)$ in Equation \eqref{fullmodel}.  Each vertex is assigned a unique community label at the outset so that the network consists of $p=|\mathbf{V}|$ communities with initial vertex community assignment $\mathbf{x}_0=\{0,1,2\ldots,p-1\}$.  The vertex labels are then iteratively resampled according to the multinomial probabilities in Equation \eqref{multinom} over $T$ full cycles of the $p=|\mathbf{V}|$ vertices and the resampled vertex labels are returned in matrix $\mathbf{X}$.  It is important to note that the Markov chain whose current state is
\begin{equation}
\centering
\begin{minipage}[][][b]{.1\textwidth}
	\includegraphics[scale=.5]{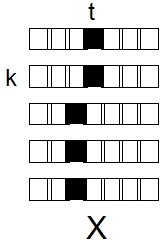}
\end{minipage}
\begin{minipage}{.8\textwidth}
\begin{equation}
\mathbf{x}^{t,k}=\nonumber
\left(\mathbf{X}[1,t],\ldots,\mathbf{X}[k,t],\mathbf{X}[k+1,t-1],\ldots,\mathbf{X}[p,t-1]\right)
\vspace{5mm}
\end{equation}
\end{minipage}
\end{equation}
for inner and outer loop iterators $k\in\{1,\ldots,p\}$ and $t\in\{2,\ldots,T\}$ of Algorithm \ref{alg1}, respectively, is not reversible.  
For example, if $v_k\in\mathbf{V}$ is isolated in its community labeled as $x_{v_k}$ and is reassigned elsewhere during resampling, the label $x_{v_k}$ becomes extinct, never to be assigned again.  This process, of course, results in a non-increasing number of communities throughout execution of Algorithm \ref{alg1}.  We have specified a cooling schedule \citep{Nourani1998} so that, consistent with standard simulated annealing procedures on Boltzmann distributions, the inverse temperature  $\theta$ may increase throughout execution to favor the resampling of higher-probability labels, see Section \ref{model}. In the hospital network application of Section \ref{appl}, we increase $\theta$ throughout execution of Algorithm \ref{alg1} so that the resampling of vertex community assignments are increasingly more greedily made.

The number of full cycles $T$ should be chosen large to ensure that the chain $\mathbf{x}^{k,t}$ converges to a local maximum of $\mathbb{P}^*_{\tau,\lambda,\theta}(\mathbf{x}|G)$ in Equation \eqref{fullmodel}, the feasibility of which we assert for $\lambda\geq1$ (see Equation \ref{optham}).  One could encode a halting criterion which checks for $\mathbf{x}^{k,t}$ arrival at a local maximum, but we have encoded an unconditional loop for simplicity of demonstration.  Once $\mathbf{x}^{k,t}$ achieves local optimality, there is no individual vertex label that can be reassigned with an increase of $\mathbb{P}_{\tau,\lambda,\theta}(\mathbf{x}|G)$.  If several vertex labels were reassigned simultaneously as opposed to one-at-a-time, however, community assignments of greater probability may be identified.  Based on this observation, the authors of the Louvain method \citep{Blondel2008} introduced a folding procedure for consolidating entire communities into individual vertices that may then be optimally relabeled using Algorithm \ref{alg1}.  We discuss the folding procedure for preparing a network-graph with a locally-optimized community assignment for further optimization in the following section.

\subsection{Folding Procedure}\label{foldproc}

The folding procedure collapses communities into vertices to facilitate vertex label reassignment at the community level and is described in Algorithm \ref{alg2}.
\afterpage{\IncMargin{1em}
\begin{algorithm}[htpb]
\spacingset{1}
\SetKwData{Left}{left}\SetKwData{This}{this}\SetKwData{Up}{up}
\SetKwFunction{Union}{Union}\SetKwFunction{FindCompress}{FindCompress}
\SetKwInOut{Input}{input}\SetKwInOut{Output}{output}
\Input{$G=(\mathbf{V},\mathbf{E}_W)$: Weighted network-graph\\
$\mathbf{x}$: Community assignment}
\Output{$G^{fold}=(\mathbf{V}^{fold},\mathbf{E}_{W^{fold}})$:}
\BlankLine
$\mathbf{U}\leftarrow\mbox{unique}(\mathbf{x})$\tcp*{unique community labels}
$\mathbf{W}^{fold}\leftarrow\mbox{matrix}(0,|\mathbf{U}|,|\mathbf{U}|)$\tcp*{an $|\mathbf{U}|\times|\mathbf{U}|$ matrix of zeros\\initialize weighted adjacency\\matrix of folded network}
\For{$i\leftarrow 1$ \KwTo $|\mathbf{U}|$}{
$\mathbf{I}\leftarrow\mbox{which}(\mathbf{x}=\mathbf{U}[i])$\tcp*{vertex indices labeled $\mathbf{U}[i]$}
\For{$j\leftarrow i$ \KwTo $|\mathbf{U}|$}{
	$\mathbf{J}\leftarrow\mbox{which}(\mathbf{x}=\mathbf{U}[j])$\tcp*{vertex indices labeled $\mathbf{U}[j]$}
	$\mathbf{W}^{fold}[i,j]\leftarrow\mbox{sum}(\mathbf{W}[\mathbf{I},\mathbf{J}])$\tcp*{sum of edge weights}\tcp*{between $u\in\mathbf{V}_{\mathbf{I}}$ and $v\in\mathbf{V}_{\mathbf{J}}$}
	$\mathbf{W}^{fold}[j,i]\leftarrow\mathbf{W}^{fold}[i,j]$\tcp*{symmeterize}
}
}
\Return $\mbox{makeGraph}(\mathbf{W}^{fold})$\tcp*{convert $\mathbf{W}^{fold}$ to graph object}
\caption{Folding Procedure}
\label{alg2}
\end{algorithm}\DecMargin{1em}}
The diagonal elements of the weighted adjacency matrix  $\mathbf{W}^{fold}$ constructed in Algorithm \ref{alg2} correspond to the sums of \textit{within} community edge weights whereas the off-diagonal elements of $\mathbf{W}^{fold}$ correspond to the sums of \textit{between} community edge weights.  With $G^{fold}=(\mathbf{V}^{fold},\mathbf{E}_{W^{fold}})$ as the output of Algorithm \ref{alg2}, we note that the folded special vertex set $\mathbf{V}^{{\prime}^{fold}}\subseteq\mathbf{V}^{fold}$ and the associated number of ICD surgeries correspond one-to-one with the original communities which contained a special vertex and the number of ICD surgeries performed within communities, respectively.  We note that if $G$ and  $\mathbf{x}^{k,t}$ are input into Algorithm \ref{alg2} with $G^{fold}$ as output, then
\begin{equation}
\begin{array}{ccc}
	Q(\mathbf{x}^{k,t}|G) = Q(\mathbf{x}_0|G^{fold}) & \mbox{ and } & T_\tau(\mathbf{x}^{k,t}|\mathbf{V})=T_\tau(\mathbf{x}_0|\mathbf{V}^{fold}),
\end{array}
\end{equation}
where $\mathbf{x}_0=\{0,1,2\ldots,|\mathbf{V}^{fold}|-1\}$.  That is, the modularity and feasibility of the community assignment $\mathbf{x}^{k,t}$ on $G$ are equivalent to the modularity and feasibility of $G^{fold}$ with the initial community assignment $\mathbf{x}_0$ of Algorithm \ref{alg1}.  Therefore, the folding procedure of Algorithm \ref{alg2} is independent of the task in solving the CMOP of identifying feasible, maximum-modularity health care communities and, instead, serves to facilitate many local optimizations in Algorithm \ref{alg1} en route to a greater maximum. In the following section, we describe alternating Algorithms \ref{alg1} and \ref{alg2} for obtaining constraint-satisfying, high-modularity vertex community assignments.

\subsection{Complete Description of Constrained Optimization Procedure}\label{complete}

In the two preceding sections, we presented the local optimization (Algorithm \ref{alg1}) and folding procedure (Algorithm \ref{alg2}) which are to be repeatedly alternated, as described in our constrained optimization procedure in Algorithm \ref{alg3}.
\afterpage{\IncMargin{1em}
\begin{algorithm}[htpb]
\spacingset{1}
\SetKwData{Left}{left}\SetKwData{This}{this}\SetKwData{Up}{up}
\SetKwFunction{Union}{Union}\SetKwFunction{FindCompress}{FindCompress}
\SetKwInOut{Input}{input}\SetKwInOut{Output}{output}
\Input{$G=(\mathbf{V},\mathbf{E}_W)$: Weighted network-graph\\
$\lambda$: Penalty weight parameter; $\theta$: Inverse temperature parameter\\
$\tau$: ICD volume threshold\\
$T$: Number of \textit{Local Optimization} (Algorithm \ref{alg1}) iterations\\
$R$: Number of \textit{Folding Operations} (Algorithm \ref{alg2})}
\Output{$\mathbf{Y}$: matrix of community labels}
\BlankLine
$\mathbf{Y}\leftarrow\mbox{matrix}(0,|\mathbf{V}|,T*R)$\tcp*{an $|\mathbf{V}|\times (T*R)$ matrix of zeros}
\For{$r\leftarrow 1$ \KwTo $R$}{
$r_0\leftarrow (r-1)T+1$\;
$r_1\leftarrow rT$\;
$\lambda\leftarrow\mbox{PenaltySchedule}(G,\mathbf{Y}[,r_1],\lambda,r)$\tcp*{default: constant}
$\mathbf{Y}[,r_0:r_1]\leftarrow\mbox{LocalOptimization}(G,\lambda,\theta,\tau,T)$\;
$G\leftarrow\mbox{FoldingProcedure}(G,\mathbf{Y}[,r_1])$\;
}
\Return $\mathbf{Y}$\;
\caption{Constrained Optimization Procedure}
\label{alg3}
\end{algorithm}\DecMargin{1em}}
As with Algorithm \ref{alg1}, one could encode a halting criterion to establish the convergence of the alternating procedure to a globally maximal vertex assignment.  The matrix $\mathbf{Y}$ returned by Algorithm \ref{alg3} contains the entire sample path of the chain $\mathbf{x}^{k,t,r}$ where
\begin{equation}\label{finalchain}
\mathbf{x}^{k,t,r} = \left(\mathbf{Y}[1,q],\ldots,\mathbf{Y}[k,q],\mathbf{Y}[k+1,q-1],\ldots,\mathbf{Y}[p,q-1]\right)
\end{equation}
and $q=(r-1)T+t$ for the inner and outer loop iterators $k\in\{1,\ldots,p\}$ and $t\in\{2,\ldots,T\}$ of Algorithm \ref{alg1}, respectively, and enveloping loop iterator $r\in\{1,2,\ldots,R\}$ of Algorithm \ref{alg3}.

\underline{Sub-routine \textit{PenaltySchedule}.} The sub-routine \textit{PenaltySchedule} of Algorithm \ref{alg3} allows the penalty parameter $\lambda$ in Equation \eqref{multinom} to vary and, in particular, control whether or not a penalty is applied.  For the purposes of the hospital network application in Section \ref{appl}, we impose a binary penalty scheme for which $\lambda=0$ (no penalty is applied) in the early stages of execution of Algorithm \ref{alg3} and $\lambda=1$ (the constraint is enforced) in later stages.  We demonstrate that for the hospital network, the penalty schedule which maintains $\lambda=0$ until no further optimization is possible before invoking $\lambda=1$ results in the highest-modularity, constraint-satisfying solution produced by our constrained community detection procedure.  Ultimately, by delaying the enforcement of the constraint until the later stages of Algorithm \ref{alg3}, we permit the vertex community labels to be assigned in early stages exclusively on the basis of network connectivity, i.e., modularity optimization.  With this understanding, our empirical results suggest that the constraint-satisfying, maximum-modularity vertex community assignment $\mathbf{x}^*$ in Equation \eqref{optham} of the nation-wide hospital network is a perturbation of the maximum-modularity vertex community assignment $\mathbf{x}^{mod}$ of Equation \eqref{modopt}.  However, the topology of other networks may warrant a different \textit{PenaltySchedule} and, therefore, as in Section \ref{appl}, we recommend several executions of Algorithm \ref{alg3} with various schedules.

\underline{Sub-routine \textit{CoolingSchedule}.} The sub-routine \textit{CoolingSchedule} of Algorithm \ref{alg1} has no bearing on the feasibility of a vertex assignment but, by promoting more greedy resampling of community labels, influences its modularity.  The sample chain $\mathbf{x}^{k,t,r}$ is, as discussed earlier, irreversible on account of the extinction of labels.  One may consider the extinction of a community label as the merging of two communities which, in early stages of execution of Algorithm \ref{alg3} when subsequent merges are expected, is of lesser consequence on the ultimate estimated community assignment as compared to later stages of execution when few operations remain.  That is, as suboptimal resampling of vertex labels become possibly more negatively influential, we increase $\theta$ with the sub-routine \textit{CoolingSchedule} in favor of greedy vertex relabeling.

\underline{Informed Search Paradigm.} In conclusion, Algorithm \ref{alg3} effectively identifies constraint-satisfying, high-modularity vertex community assignments.  In fact, the sample chain $\mathbf{x}^{k,t,r}$ in Equation \eqref{finalchain} produced by Algorithm \ref{alg3} may be regarded as the path of an informed search through the space of all community assignments $\mathbf{K}_p$.  Thus, we estimate $\mathbf{x}^*$ in Equation \eqref{optim} with
\begin{equation}\label{dag}
	\mathbf{x}^{\dagger} = \underset{\substack{\mathbf{x}^{r,t,r}\\1\leq k\leq p;1\leq t\leq T;1\leq r\leq R}}{\arg\max}\left\{Q(\mathbf{x}^{k,t,r}|G):\mathbf{x}^{k,t,r}\in\mathbf{S}_\tau(\mathbf{V})\right\}
\end{equation}
as the maximum-modularity, constraint-satisfying vertex assignment traversed by the MCMC procedure.  More generally, the \textit{PenaltySchedule} sub-routine in Algorithm \ref{alg3} could be set $\lambda=0$ identically so that, although there is no guarantee that the chain $\mathbf{x}^{k,t,r}$ terminates in a feasible state, it will (up to sampling variation) terminate with greater modularity than its constrained counterpart.  Thus, we estimate $\mathbf{x}^{mod}$ in Equation \eqref{modopt}
with
\begin{equation}\label{ddag}
	\mathbf{x}^{\ddagger} = \underset{\substack{\mathbf{x}^{r,t,r}\\1\leq k\leq p;1\leq t\leq T;1\leq r\leq R}}{\arg\max}\left\{Q(\mathbf{x}^{k,t,r}|G)\right\}
\end{equation}
as the maximum-modularity vertex assignment discovered by the procedure.

In the following application of our constrained community detection procedure on the nation-wide hospital network we generate several instances of the unconstrained and constrained chains and their optima $\mathbf{x}^\ddagger$ and $\mathbf{x}^\dagger$ to ultimately overcome the Monte Carlo error and select the best performing chain of each type as our estimators for $\mathbf{x}^{mod}$ and $\mathbf{x}^*$, respectively.

\subsection{Setup for the Estimation of Health Care Communities}

Our analysis of the health care community (HCC) structure of the nation-wide hospital network begins with estimating the unconstrained communities by executing Algorithm \ref{alg3} with $\lambda=0$ throughout. We utilize the unconstrained communities, in part, as a basis for comparison with the constraint-satisfying HCCs which we subsequently estimate. Additionally, we specify the minimum threshold of ICD surgery volume per HCC as $\tau=\lfloor\sum_{v\in\mathbf{V}}f(v)/|\mathbf{x}^{mod}|\rfloor$, where  $|\mathbf{x}^{mod}|$ is the number of unconstrained communities, as the average ICD surgery volume per unconstrained community in the nation-wide network.  This choice for $\tau$ is made for ease of demonstration in the present application since it is a function of only one unknown quantity: the value of $|\mathbf{x}^{mod}|$.  That is, we estimate the value $|\mathbf{x}^{mod}|$ with $|\mathbf{x}^{\ddagger}|$, where $\mathbf{x}^\ddagger$ is the estimated unconstrained maximum-modularity community assignment in Equation \eqref{ddag} and estimate subsequently the constrained communities by executing Algorithm \ref{alg3} with $\tau$ equal to the mean ICD volume per estimated unconstrained community in the estimated assignment $\mathbf{x}^\ddagger$. We illustrate the MCMC sample paths resulting from the invocation of penalization at the two intermediate folding points during optimization of $\mathbb{P}^*_{\tau,0,\theta}(\mathbf{x}|G)$ and, additionally, at the end of the chain. During both the unconstrained and constrained estimation procedures we use \textit{CoolingSchedule} to define the exponential rate $\theta= 2^{t-1}$ of inverse temperature in Algorithm \ref{alg1}, where $t\in\{2,\ldots,T\}$ is the outer-loop iterator of the procedure (see \cite{Nourani1998} for a comparison of cooling rates).  We select an exponential cooling rate based on the context of our tolerance for sub-optimal community label reassignments, as discussed in Section \ref{complete}. We compare the constrained solutions to the unconstrained solutions to identify qualitative differences between the two sets of community assignments and, crucially, we investigate how the minimum ICD volume constraint impacts community organization in the nation-wide hospital network.

\addtocounter{section}{1}
\section*{\centering \arabic{section}. \uppercase{Estimation of Health Care Communities}}
\addtocounter{section}{-1}
\refstepcounter{section}
\label{appl}

The nation-wide hospital network we consider consists of $p=4734$ hospitals, among which $1388$ (29.3\%) are CCFs hosting ICD surgeries and correspond to special vertices in $\mathbf{V}^\prime$.  The weighted degree distributions of all hospitals and the CCFs are provided in Figure \ref{one}a.
\afterpage{
\begin{figure}[!htbp]
\centering
\includegraphics[width=.49\textwidth,height=.25\paperheight]{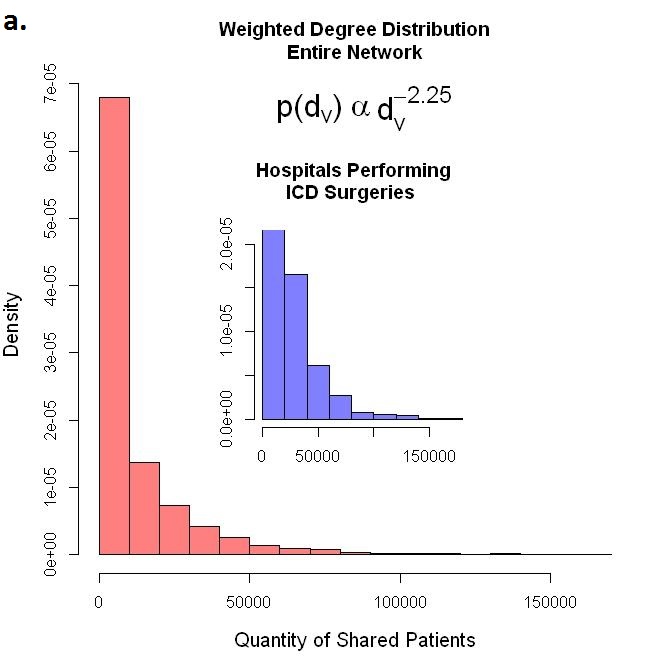}
\includegraphics[width=.49\textwidth,height=.25\paperheight]{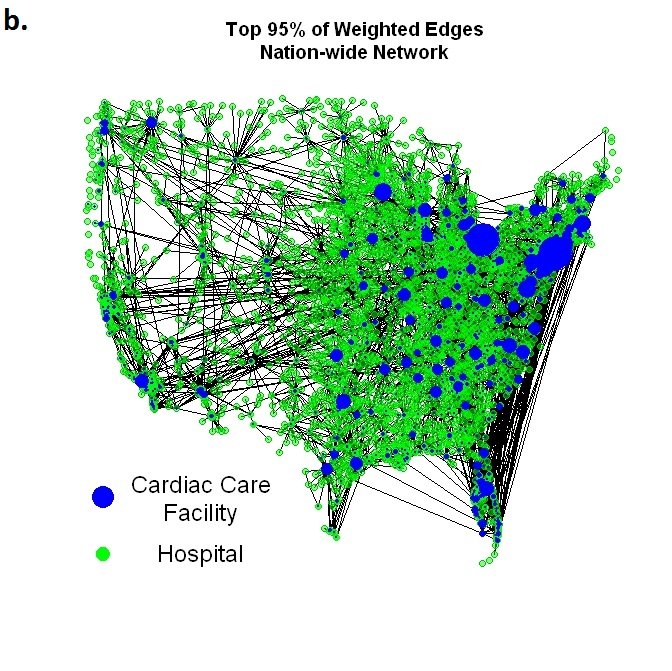}
\caption{\linespread{1}\selectfont{}Nation-wide hospital network.  \textbf{a.} Weighted degree distribution for the entire network (red) and hospital where ICD surgeries are performed (blue) with medians $3510.3$ and $22157.2$, respectively.  \textbf{b.} The hospitals where ICD are performed are predominantly located in Eastern states and notably in the cities of New York, NY and Cleveland, OH.}
\label{one}
\end{figure}}
The United States map in Figure \ref{one}b indicates that the bulk of ICD surgeries occur in hospitals located in the Eastern States and have a higher frequency of shared patients with physicians associated with other hospitals compared to the entire population of hospitals in the network.

\subsection{Unconstrained Hospital Network Communities}\label{uncond}

To estimate unconstrained communities in the hospital network, we set the \textit{PenaltySchedule} sub-routine to return $\lambda=0$ (so that no penalty is applied) in Algorithm \ref{alg3} to estimate $\mathbf{x}^{mod}$ of Equation \eqref{modopt}.  Empirical results from $250$ independent unconstrained sample paths $\mathbf{x}^{k,t,r}_\ddagger$, where the subscript $\ddagger$ indicates an unconstrained chain (see Equation \ref{ddag}), generated by Algorithm \ref{alg3} for the nation-wide hospital network are displayed in Figure \ref{two}a.
\afterpage{
\begin{figure}[!htbp]
\centering
\includegraphics[width=.32\textwidth,height=.2\paperheight]{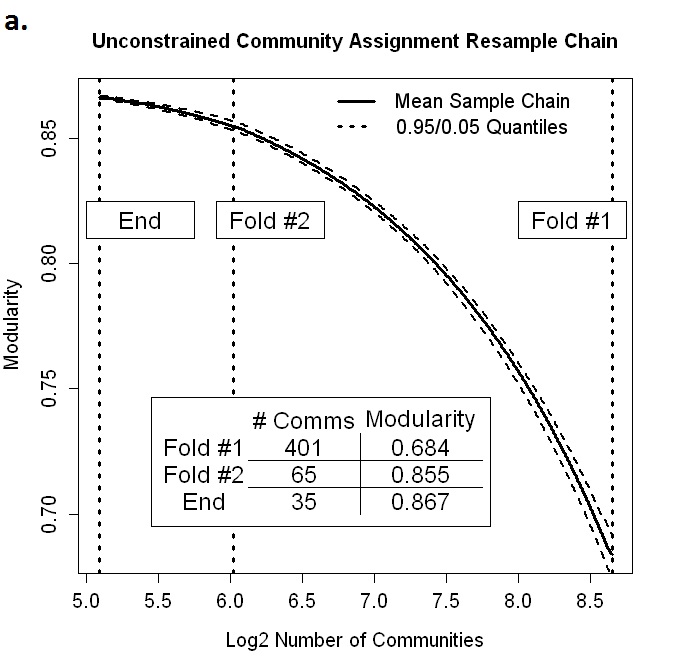}
\includegraphics[width=.32\textwidth,height=.2\paperheight]{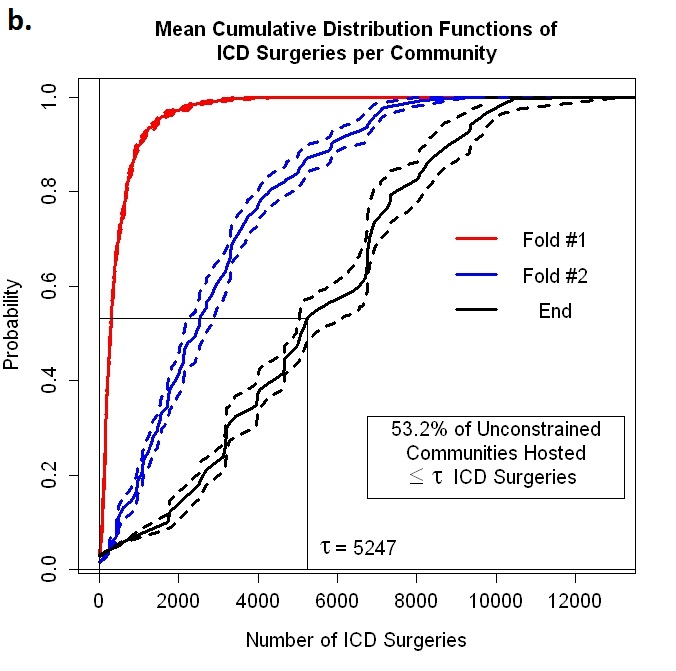}

\includegraphics[width=.32\textwidth,height=.2\paperheight]{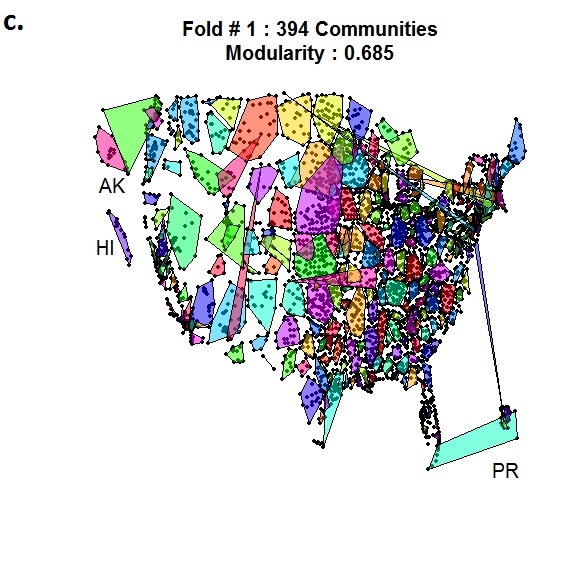}
\includegraphics[width=.32\textwidth,height=.2\paperheight]{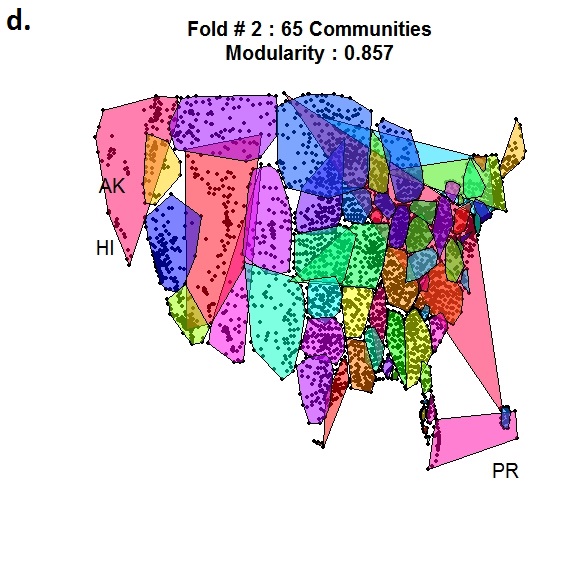}
\includegraphics[width=.32\textwidth,height=.2\paperheight]{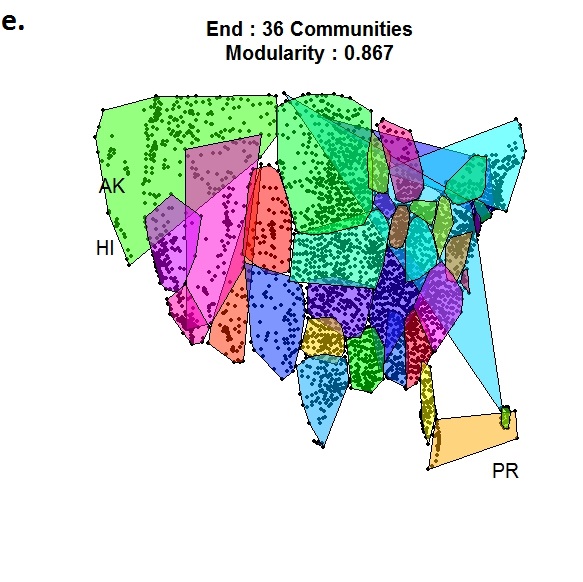}
\caption{\linespread{1}\selectfont{}\textbf{a.} The average modularity of $250$ unconstrained sample paths $\mathbf{x}_\ddagger^{k,t,r}$. \textbf{b.} The average cumulative distribution functions of ICD surgery volume per unconstrained community at each local max / folding point. Upon termination of the procedure, $52.8\%$ of unconstrained communities have an ICD surgery volume at most $\tau=5247$. \textbf{c.-e.} United States map with unconstrained communities at each local maximum / folding point indicated by polygons.  The number of communities decreases as modularity increases during the course of execution of Algorithm \ref{alg3} and, at its termination (\textbf{e.}) there are $36$ communities in $\mathbf{x}^\ddagger$ for a modularity of $0.867$.
}
\vspace{-10pt}
\label{two}
\end{figure}}
The mean sample path $\mathbf{\bar{x}}_\ddagger^{k,t,r}$ traces a path through $\mathbf{K}_p$ with a trend of increasing modularity that, up to sampling variation, is necessarily an upper-bound for any penalized sample chain.  Each path $\mathbf{x}_\ddagger^{k,t,r}$ proceeds from right-to-left as a consequence of the agglomerative vertex label resampling process in which community labels may become extinct during optimization, see Section \ref{locopt}. The vertical lines indicate the points at which the average sample path achieves a local optima of $\mathbb{P}^*_{\tau,0,\theta}(\mathbf{x}|G)$ (Algorithm \ref{alg1}) and the folding mechanism (Algorithm \ref{alg2}) was initiated.  Among the $250$ instances of $\mathbf{x}_\ddagger^{k,t,r}$, the estimated community assignments obtained a maximum modularity of $0.867$.

The folding points in the unconstrained optimization procedure are natural break points for invoking the penalty with \textit{PenaltySchedule} in Algorithm \ref{alg3} to enforce constraint satisfaction, as is done in the following section.  A map of the United States with overlaying polygons representing the communities discovered at each of these local maxima / folding points of the unconstrained sample path $\mathbf{x}^{k,t,r}_\ddagger$ are presented in Figures \ref{two}c-\ref{two}e  We illustrate the constrained sample chains $\mathbf{x}^{k,t,r}_\dagger$ as they originate from these three breaks points in the sample chain produced by the optimization process of Algorithm \ref{alg3} in the following.

\subsection{Heath Care Communities}

We now estimate the health care community assignment $\mathbf{x}^*$ of Equation \eqref{optham} that optimizes modularity subject to the constraint that the total volume of ICD surgeries performed within each HCC exceeds the threshold $\tau$, as prescribed in the formal statement of the CMOP.  We set $\tau=\lfloor\sum_{v\in\mathbf{V}}f(v)/|\mathbf{x}^\ddagger|\rfloor=5247$ as the average number of ICD surgeries per estimated unconstrained community in the assignment $\mathbf{x}^\ddagger$ of Equation \eqref{ddag}.
\afterpage{
\begin{figure}[!htbp]
	\centering
	\includegraphics[width=.32\textwidth,height=.2\paperheight]{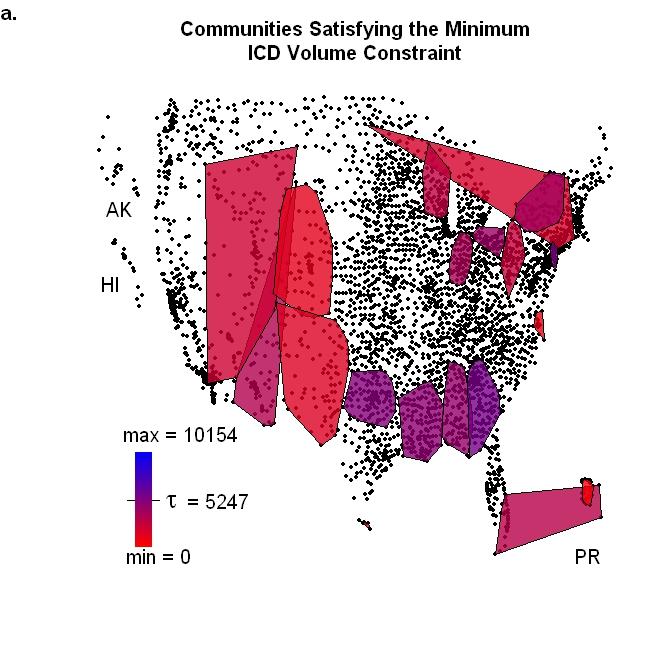}
	\includegraphics[width=.32\textwidth,height=.2\paperheight]{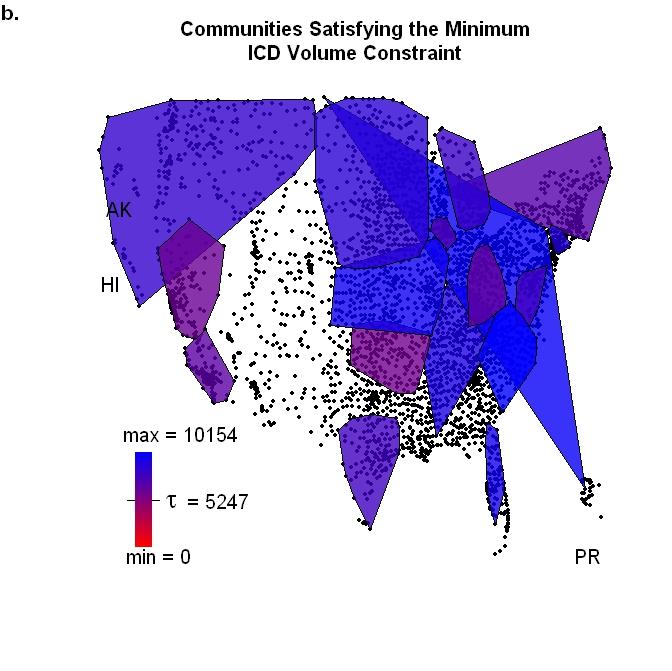}
	\includegraphics[width=.32\textwidth,height=.2\paperheight]{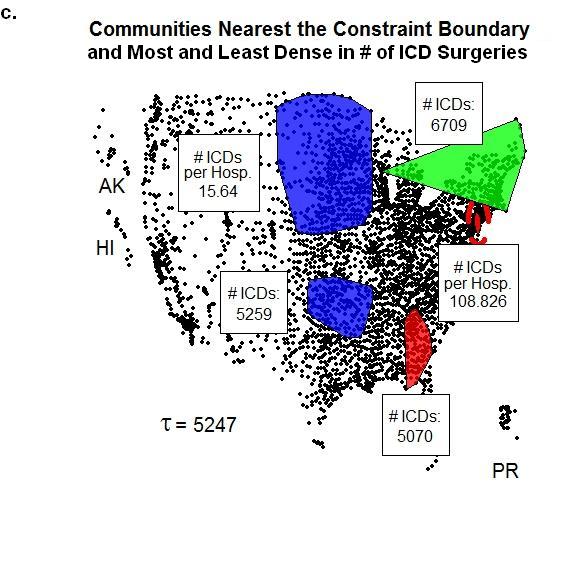}
	
	\includegraphics[width=.32\textwidth,height=.2\paperheight]{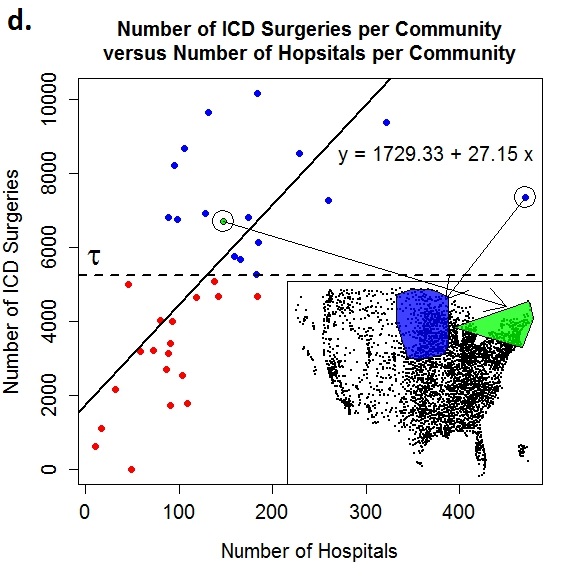}
	\includegraphics[width=.32\textwidth,height=.2\paperheight]{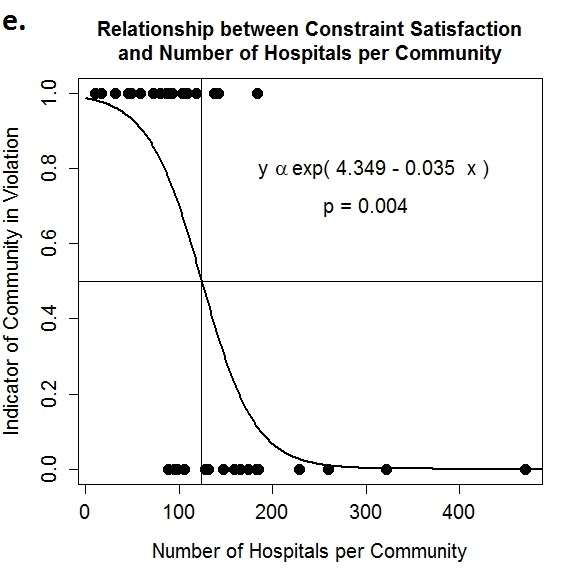}
	\includegraphics[width=.32\textwidth,height=.2\paperheight]{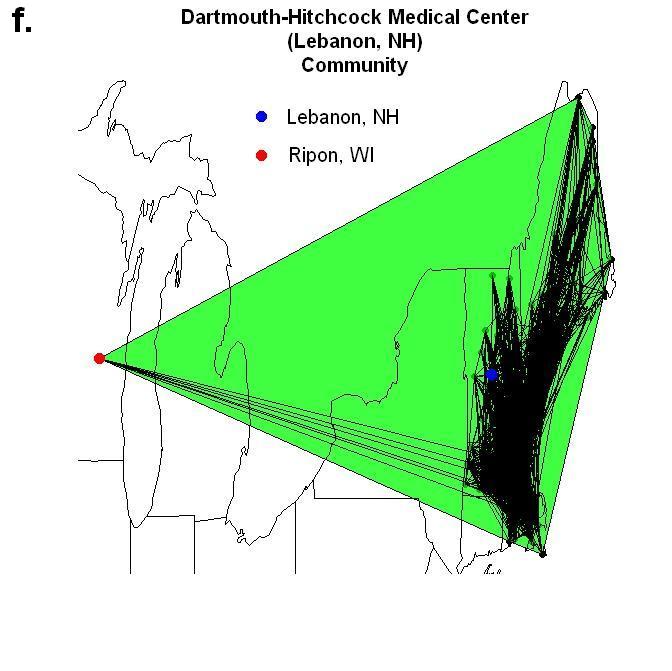}
	\caption{\linespread{1}\selectfont{}Depiction of unconstrained communities.  \textbf{a}. Communities with an insufficient ICD volume, i.e., in violation of the constraint.  \textbf{b}.  Communities with a sufficient ICD surgery volume, i.e., the constraint is satisfied. \textbf{c}.  The blue communities are constraint-satisfying whereas the red communities are not.  The least- and most-dense communities that satisfy and do not satisfy the constraint, respectively, are plotted along with the communities nearest the constraint boundary.  Dartmouth-Hitchcock Medical Center (DH) belongs to the New England community that contains one hospital at a distance from Ripon, WI.  The DH community is in compliance with the constraint. \textbf{d}.  The increasing trend between the number of ICD surgeries performed at hospitals within a community and the number of hospitals within a community.  The community representing the most hospitals is also the least dense. The DH community is near the trend line. \textbf{e.} Logistic regression of the infeasibility indicator variable $T_\tau(\mathbf{x},\mathbf{V})$ on the number of hospitals per community. Communities containing more hospitals are more likely to be constraint-satisfying.  \textbf{f.}  Edges within the DH community.  The hospital in Ripon, WI has several shared patients with hospitals in Massachusetts.  The unconstrained community was shown in green throughout panels c., d., and f.}
	\label{three}
\end{figure}}

\underline{(In)feasibility of Unconstrained Communities.} The maps in Figures \ref{three}a and \ref{three}b depict the unconstrained communities that are in violation of and those that satisfy the constraint, respectively.  The geographic dispersion of the communities in violation are among the three main regions of the Western Mountain and Desert region, Southern region, and the Lower Great Lakes region.  The community isolated on  Puerto Rico is most in violation with zero ICD surgeries taking place at hospitals assigned to it, e.g., there are no CCFs belonging to the island community.  

The map in Figure \ref{three}c similarly depicts the unconstrained communities with a number of ICD surgeries nearest the threshold $\tau=5247$ and, additionally, the most dense and least dense communities that are in violation and which satisfy the constraint, respectively.  The community to which Dartmouth-Hitchcock Medical Center (DH) belongs is situated in Lebanon, NH \if0\blind{(location of the authors)}\fi\hspace{0pt} and covers most of New England while incorporating one hospital at a distance.  This hospital is located in Ripon, WI and is well-connected to the hospitals belonging to the DH community, see Figure \ref{three}f.  While DH contributed a volume of $f(v_{DH})=461$ ICD surgeries toward the $F(v_{DH},\mathbf{x}^\ddagger)=6709$ of its community total, the hospital in Ripon, WI is not a CCF and thus did not contribute any volume of ICD surgeries.  We note that facilitating the grouping together of hospitals of vast geographic distances is characteristic of our network-based approach to allocating nation-wide health services that was previously impossible in the geographically-defined HRRs.

\underline{Constrained Community Detection.} We employ the constrained community detection procedure described in Algorithm \ref{alg3} to identify constraint-satisfying communities by invoking the penalty terms of $\mathbb{P}^*_{\tau,\lambda,\theta}(\mathbf{x}|G)$ (Equation \ref{fullmodel})  with the sub-routine \textit{PenaltySchedule} during optimization.  In Figure \ref{foura}a., we illustrate the sample chains $\mathbf{x}_{\dagger j}^{k,t,r}$, when the $\lambda$ is switched from $0$ to $1$ at folding points $j=1,2,3$, where $j=3$ corresponds to the end of the unconstrained chain $\mathbf{x}_\ddagger^{k,t,r}$.  The subscript $\dagger$ in $\mathbf{x}_{\dagger j}^{k,t,r}$ indicates a constrained chain (see Equation \ref{dag}).
\afterpage{
\begin{figure}[!htbp]
\centering
\includegraphics[width=.32\textwidth,height=.2\paperheight]{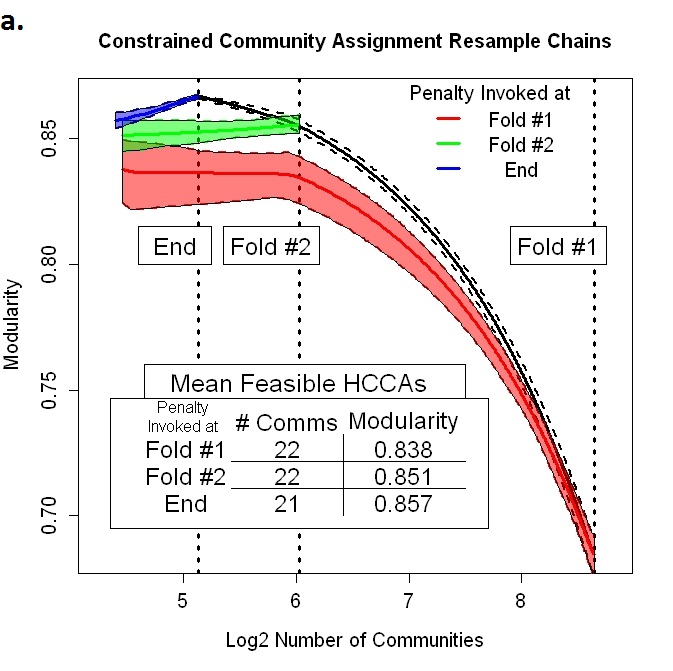}
\includegraphics[width=.32\textwidth,height=.2\paperheight]{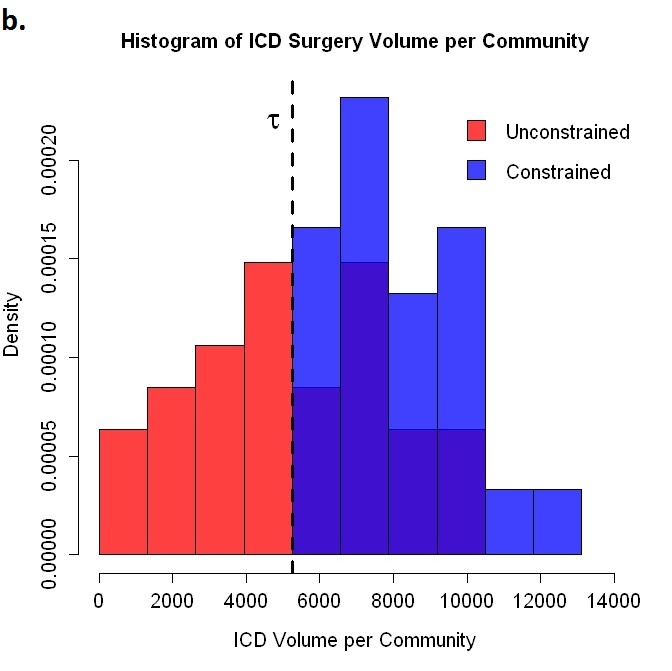}

\includegraphics[width=.32\textwidth,height=.2\paperheight]{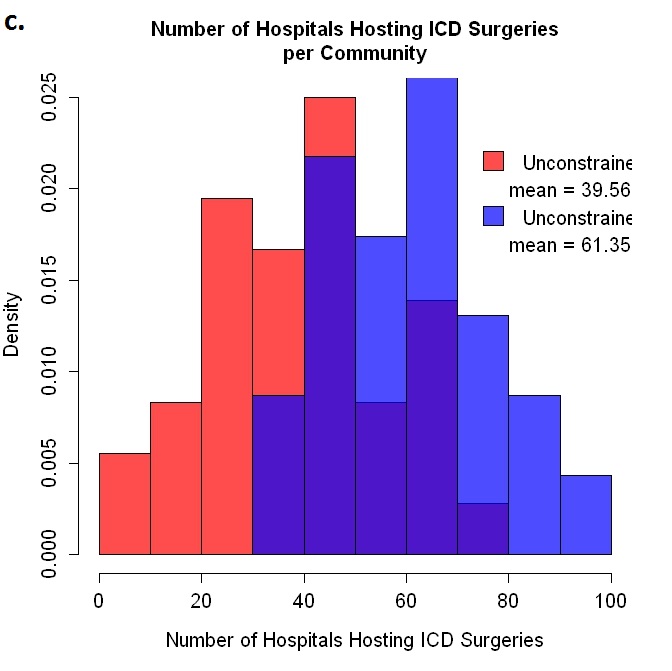}
\includegraphics[width=.32\textwidth,height=.2\paperheight]{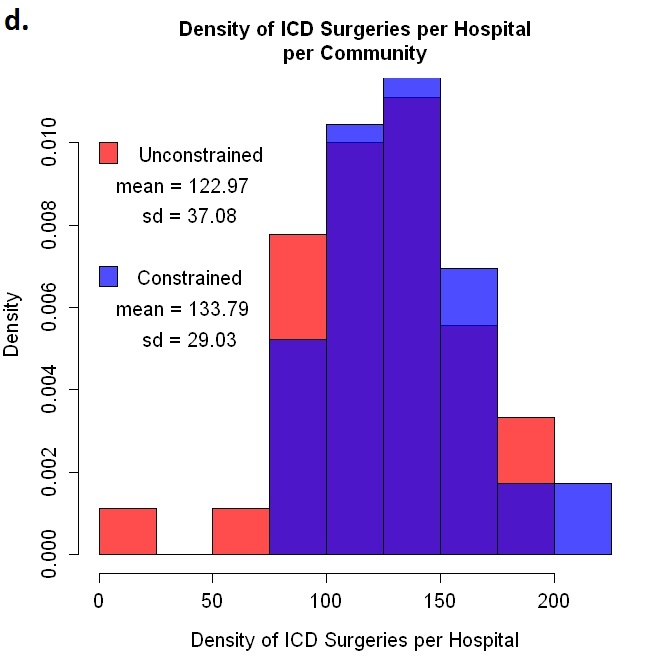}
\caption{\linespread{1}\selectfont{}Comparison between unconstrained and constrained communities.  \textbf{a.} MCMC sample paths (mean and 90\% CI) for unconstrained optimization (black) and constrained optimization initiated at Fold \#1 (red), \#2 (green), and the end (\#3) of the unconstrained optimization (blue).  The vertical dashes indicated the folding points of the unconstrained optimization procedure. \textbf{b.} The estimated HCC assignment $\mathbf{x}^\dagger$ specifies communities that contain an ICD surgery volume exceeding $\tau=5247$.  \textbf{c.}  The number of CCFs (hospitals hosting ICD surgeries) per community is approximately $21.79$ greater in the HCC assignment $\mathbf{x}^\dagger$ compared to its unconstrained counterpart $\mathbf{x}^\dagger$.  \textbf{d.} The ratio of ICD surgeries per hospital is similarly distributed among communities between assignments $\mathbf{x}^\dagger$ and $\mathbf{x}^\ddagger$.}
\label{foura}
\end{figure}}

\afterpage{
\begin{figure}[!htbp]
\centering
\includegraphics[width=.32\textwidth,height=.2\paperheight]{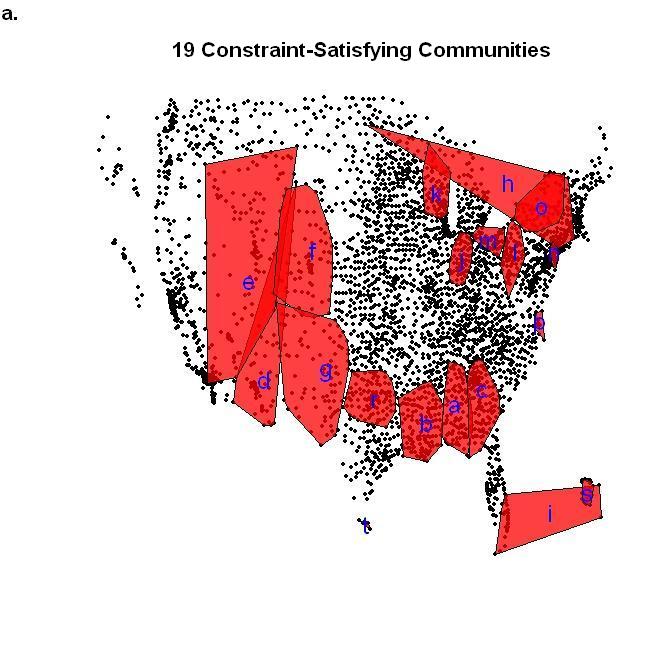}
\includegraphics[width=.32\textwidth,height=.2\paperheight]{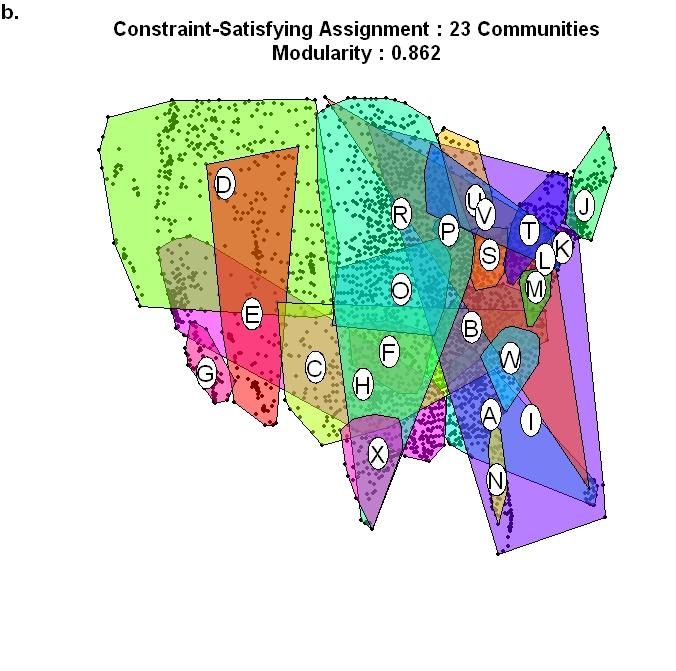}
\includegraphics[width=.32\textwidth,height=.2\paperheight]{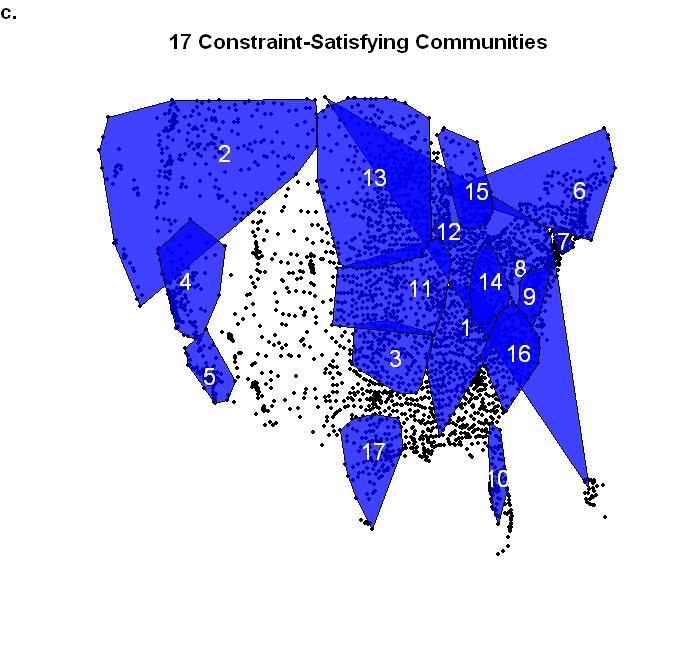}

\includegraphics[width=.32\textwidth,height=.2\paperheight]{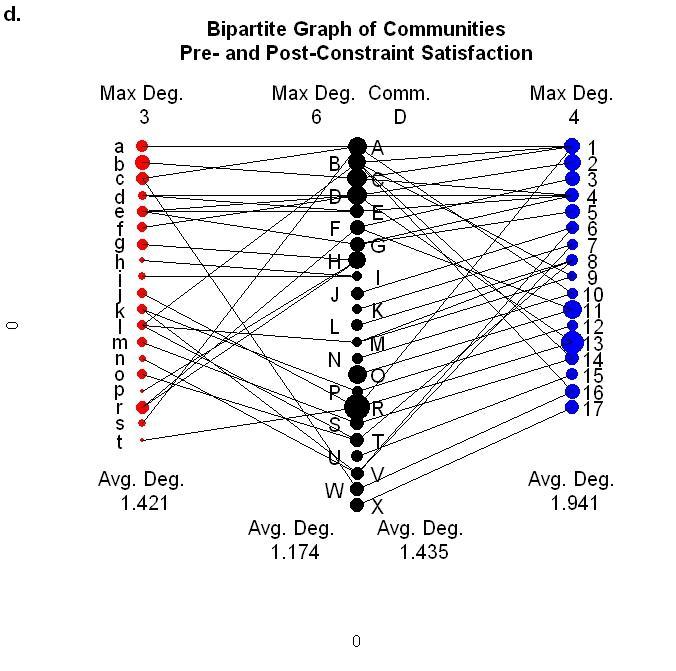}
\includegraphics[width=.32\textwidth,height=.2\paperheight]{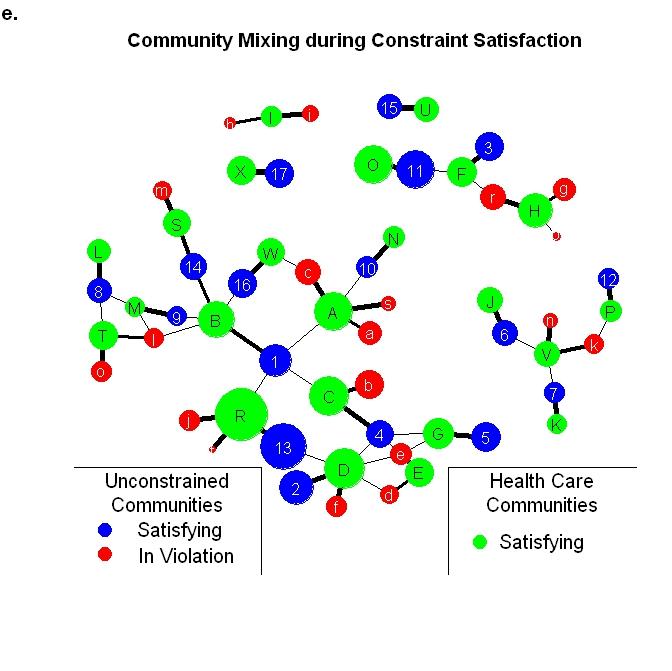}
\includegraphics[width=.32\textwidth,height=.2\paperheight]{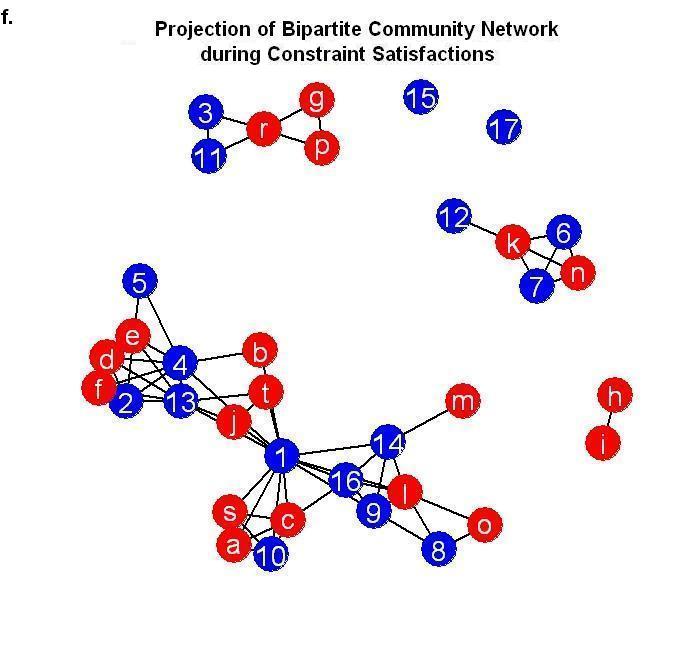}
\caption{\linespread{1}\selectfont{}Reconfiguration of communities during constraint satisfaction.  \textbf{a.} Unconstrained communities in violation of the minimum ICD volume constraint.  Each is identified by a unique lower-case letter.  \textbf{b.} Health care communities.  Each is identified by a unique upper-case letter.  \textbf{c.} Unconstrained communities satisfying the constraint.  Each is identified by a unique number.  \textbf{d.} The red and blue communities in a. and c. are identified by the community assignment $\mathbf{x}^\dagger$ from which the constrained optimization is initialized.  The manner in which these communities are related to the HCCs subsequently estimated (according to the vertices incorporated) is depicted in the bipartite graph. \textbf{e.} The graph of unconstrained communities (red and blue) along with HCCs (green).  Vertex diameter is proportional to the number of hospitals within the community and edge width is proportional to the number of hospitals in common. \textbf{f.} Most unconstrained communities are affiliated through vertex-sharing to other communities except blue communities $15$ and $17$, which remained intact throughout the constrained optimization procedure.  Additionally, red communities \textit{i} and \textit{h} merged to form a constraint-satisfying HCC.}
\label{fourb}
\end{figure}}

Each of the $750$ constrained chains (250 for each $j=1,2,3$) is, upon termination but possibly earlier, a feasible community assignment in the space $\mathbf{S}_\tau(\mathbf{V})$.  Among the three types of MCMC paths $\mathbf{x}_{\dagger j}^{k,t,r}$, each initiated at the $j^{th}$ folding point, for $j=1,2,3$, the chain $\mathbf{x}_{\dagger 3}^{k,t,r}$ in which the penalty was invoked only after the unconstrained chain $\mathbf{x}_\ddagger^{k,t,r}$ had terminated produced the highest-modularity community assignment both on average and at its maximum, see Figure \ref{foura}a.  Likely attributable to the delayed onset of constraint enforcement, the median number of communities per assignment among these higher-modularity community assignments $\mathbf{x}^{k,t,r}_{\dagger 3}$ is one fewer than their earlier-initiated counterparts $\mathbf{x}^{k,t,r}_{\dagger j}$, for $j=1,2$.  That is, by the time the penalty was invoked on the chain $\mathbf{x}_{\ddagger 3}^{k,t,r}$, the high-modularity community structure is well-established, as opposed to when the penalty was invoked earlier in the chains $\mathbf{x}_{\dagger j}^{k,t,r}$, for $j=1,2$.  As a result, the vertex community labels are less rigid and more readily modified in Algorithm \ref{alg3}, see Section \ref{complete} for a discussion.  In accordance with Equation \eqref{dag}, we identify the maximum-modularity feasible assignment over all $750$ chains as
\begin{equation}\label{fdag}
\mathbf{x}^{\dagger} = \underset{\substack{\mathbf{x}_{\dagger j}^{r,t,r}\\1\leq k\leq p;1\leq t\leq T;1\leq r\leq R\\j=1,2,3}}{\arg\max}\left\{Q(\mathbf{x}_{\dagger j}^{k,t,r}|G):\mathbf{x}_{\dagger j}^{k,t,r}\in\mathbf{S}_\tau(\mathbf{V})\right\}.
\end{equation}

The modularity $Q(\mathbf{x}^\dagger|G)=0.8615$ of the constrained maximum-modularity community assignment $\mathbf{x}^{\dagger}$ is a $0.0052$ reduction from the unconstrained maximum-modularity community assignment $\mathbf{x}^\dagger$ of $Q(\mathbf{x}^\ddagger)=0.8667$.  The estimated health care community assignment $\mathbf{x}^\dagger$ in Equation \eqref{fdag} designates $|\mathbf{x}^\dagger|=23$ communities compared with an estimated $|\mathbf{x}^\ddagger|=36$ unconstrained communities.

The histogram in Figure \ref{foura}b illustrates the rightward shift in ICD surgery volume per community and, crucially, that each of the HCCs (constrained communities) in the assignment $\mathbf{x}^\dagger$ exceeds the minimum threshold $\tau=5247$.  Interestingly, the densities of ICD surgeries per hospital per community are similarly distributed across both community assignments $\mathbf{x}^\ddagger$ and $\mathbf{x}^\dagger$.
The unconstrained communities with an ICD surgery volume at most $\tau$ are depicted in red in Figure \ref{fourb}a. and are assigned an identifying lowercase letter whereas the unconstrained communities with ICD surgery volume exceeding $\tau$ are depicted in blue in Figure \ref{fourb}c and are assigned an identifying number (these are the same communities identified in Figure \ref{three}a and \ref{three}b).  The community structure corresponding to the constraint-satisfying  vertex assignment $\mathbf{x}^\dagger$ is depicted in Figure \ref{fourb}b.

A noteworthy distinction between the unconstrained and constrained community structure is the Southern Florida and Puerto Rico community \textit{i} (Figure \ref{fourb}a), which now belongs to community \textit{I} (Figure \ref{fourb}{b}), which spans a far greater geographic region than before to include some hospitals as far north as New York -- a possible consequence of the so-called ``snowbird'' effect \citep{McHugh1994} from the seasonal migration of Medicare enrollees.  That is, Southern Florida may be a well-connected and high-modularity sub-network in isolation but, when the ICD volume constraint was applied, the CCFs in New York where, presumably, many of the seasonal migrants receive cardiac care merged with the regionally distant Southern Florida community.  Additionally, community $6$ to which DH belongs no longer includes the hospital in Ripon, WI once the constraint has been satisfied.  The Ripon, WI hospital was assigned during the constrained optimization procedure to the more geographically proximal community \textit{k} which primarily merged with community \textit{n} to form feasible community \textit{V}.  Since the Ripon, WI hospital was not contributing any ICD surgeries to its unconstrained community volume (which exceeded $\tau$) its departure must be on account of an increase in the interaction component $\mathcal{H}^I(\mathbf{x}|G)$ of $\mathbb{P}^*_{\tau,\lambda,\theta}(\mathbf{x}|G)$ due to joining with community \textit{k}.  Otherwise, the New England community $6$ remains relatively geographically stable as community \textit{J} in the feasible assignment $\mathbf{x}^\dagger$.

The unconstrained communities, both those in compliance with the constraint and those violating it, merge and exchange hospitals during the process of adapting the estimated maximum-modularity vertex assignment $\mathbf{x}^\ddagger$ to the estimated health care community assignment $\mathbf{x}^\dagger$.  Note in Figure \ref{fourb}e that unconstrained communities $h$ and $i$ are both in violation of the constraint and, during the constrained phase of Algorithm \ref{alg3}, the two communities merged to form the health care community \textit{I} that satisfies the constraint.  Only the two unconstrained communities $15$ and $17$, located in Michigan and Texas, were initially feasible and were unmodified during constrained optimization.  Each unconstrained community in violation of the constraint contributed vertices to an average of $1.421$ constrained communities whereas unconstrained communities satisfying the constraint contributed vertices to an average of $1.941$ constrained communities. The constrained community associated with the greatest number of unconstrained communities, drawing vertices from six unconstrained communities, is identified as \textit{D} and is located in the northwestern United States.

We identify all vertex pairs that were assigned to the same community in $\mathbf{x}^\ddagger$, $\mathbf{x}^\dagger$, or both and display the fraction of each type in Figure \ref{six}a.  By a factor of nearly two-to-one, vertex pairs fitting this description were assigned to the same community in both $\mathbf{x}^\ddagger$ and $\mathbf{x}^\dagger$.  Far less frequently occurring were vertex pairs belonging to the same constrained community that were separated during constrained optimization.  The remaining one-third of vertex pairs that were at some point assigned to the same community, were assigned similarly only in the feasible community assignment $\mathbf{x}^\dagger$ -- primarily attributable to the merging of communities during the constraint-satisfaction procedure.  We note that this trend is reversed among vertex pairs for which at least one belongs to an unconstrained community satisfying the constraint.
\afterpage{
\begin{figure}[!htbp]
\centering
\includegraphics[width=.32\textwidth,height=.2\paperheight]{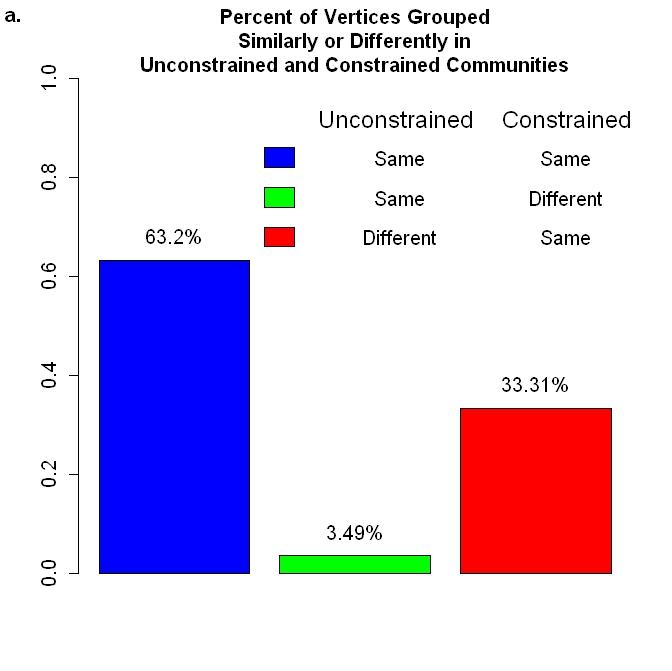}
\includegraphics[width=.32\textwidth,height=.2\paperheight]{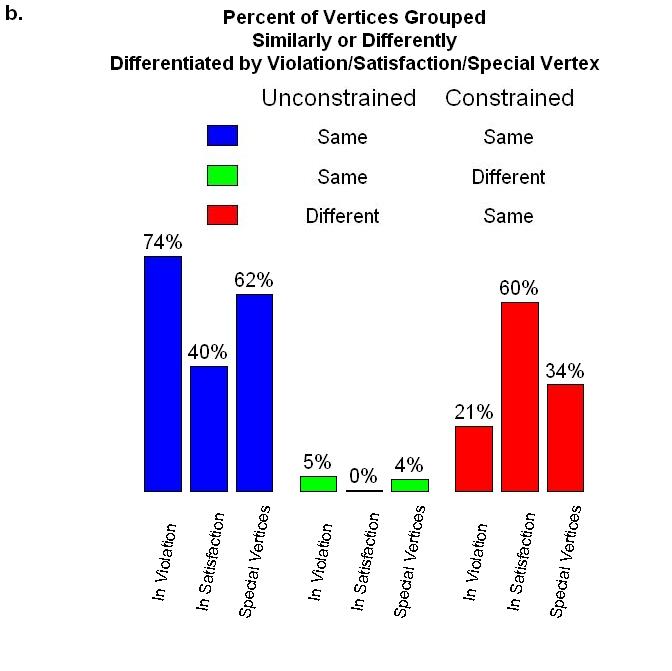}
\includegraphics[width=.32\textwidth,height=.2\paperheight]{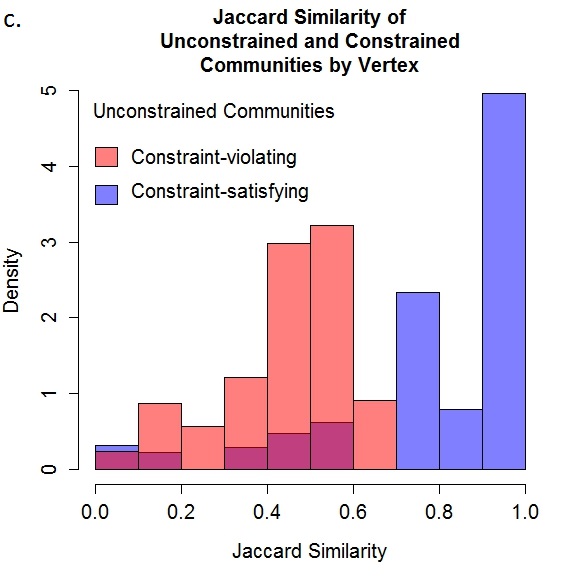}

\includegraphics[width=.32\textwidth,height=.2\paperheight]{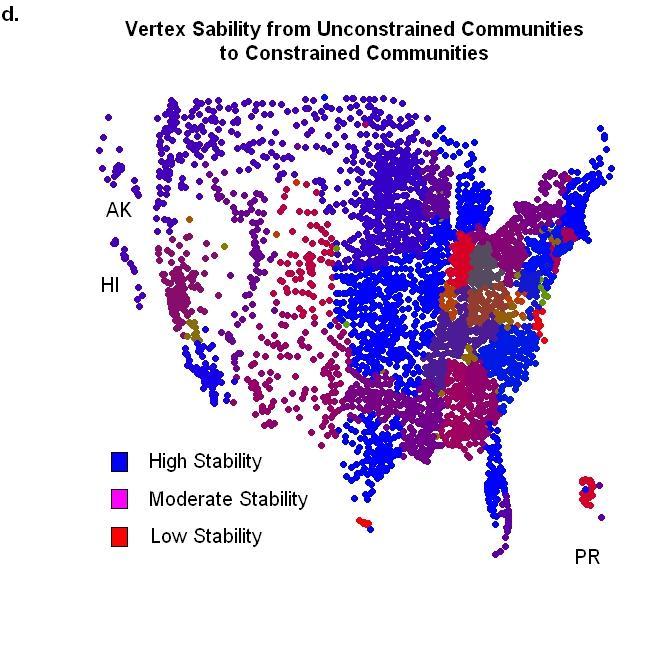}
\includegraphics[width=.32\textwidth,height=.2\paperheight]{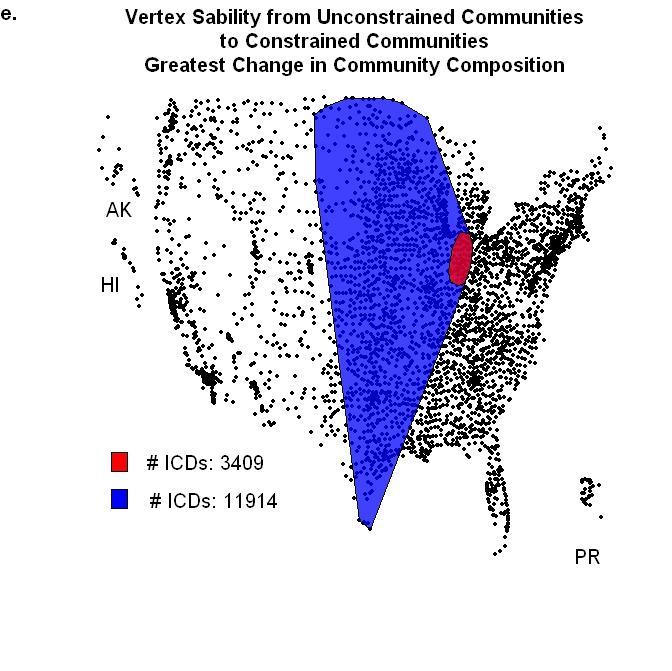}
\caption{\linespread{1}\selectfont{}\textbf{a.} Among vertex pairs for which both vertices belong to the same unconstrained community or the same HCC after constrained optimization, $63.2\%$ were in the same community pre- and post-optimization.  The $33.31\%$ of vertex pairs were labeled differently pre-optimization and the same post-optimization.  \textbf{b.} Differentiated by vertex type, a difference between vertex pairs labeled differently pre- but the same post-constrained optimization are nearly three-times more likely to have at least one vertex belonging to a community satisfying the constraint pre-constrained optimization.  \textbf{c.} Jaccard similarity (see text) of the unconstrained and constrained communities to which an individual vertex belongs. \textbf{d.} The color of each vertex is on the RGB scale corresponding to the colors in a.  A vertex belonging to an unconstrained community with many of the same (different) vertices in its HCC is stable (unstable) and of blue (red) hue.  \textbf{e.} The red community geographically corresponds to Indiana and, in addition to being in violation of the constraint, is considered unstable.  Post-constrained optimization, the hospitals in Indiana are grouped with the largest unconstrained community which satisfies the constraint and is located in the Upper-Midwest.  The small community located in Southern Texas is similarly grouped.}
\label{six}
\end{figure}}

On an individual vertex basis, as opposed to a vertex-pair basis, vertices which belong to a feasible unconstrained community are significantly more likely (see Figure \ref{six}c) to be associated with the same set of vertices after the constrained optimization procedure is complete.  For an individual vertex $v\in\mathbf{V}$, the Jaccard measure of similarity $J_v$ between the unconstrained and constrained communities $\mathbf{U}^{\ddagger}_v,\mathbf{U}_v^{\dagger}\subseteq\mathbf{V}$, respectively, is defined as
\begin{equation}
	J_v = \frac{|\mathbf{U}^\ddagger_v\cap\mathbf{U}^\dagger_v|}{|\mathbf{U}^\ddagger_v\cup\mathbf{U}^\dagger_v|}.
\end{equation}
The value $J_v$ evaluates the ratio of the number of vertices belonging to both $\mathbf{U}^\ddagger_v$ and $\mathbf{U}^\dagger_v$ to the number of vertices belonging to at least one of $\mathbf{U}^\ddagger_v$ or $\mathbf{U}^\dagger_v$.  This result is, in part, explained by the fact that vertices belonging to unconstrained communities in violation of the constraint end up belonging to constrained communities that contain, on average, 112.52 more vertices than constrained communities to which vertices belonging to unconstrained communities that satisfy the constraint belong.  That is,
\begin{eqnarray*}
\mathbb{E}\left[|\mathbf{U}_v^\ddagger|-|\mathbf{U}_v^\dagger||F(v,\mathbf{x}^\ddagger)>\tau\right] &=& -45.503 \\ \mathbb{E}\left[|\mathbf{U}_v^\ddagger|-|\mathbf{U}_v^\dagger||F(v,\mathbf{x}^\ddagger)\leq\tau\right] &=& -158.018,
\end{eqnarray*}
where, as a basic consequence of fewer constrained than unconstrained communities, both quantities are negative.

We plot in Figure \ref{six}d the stability of a vertex within its community assignment in moving from $\mathbf{x}^\ddagger$ to $\mathbf{x}^\dagger$.  Each vertex $v\in\mathbf{V}$ has an associated vertex subset $\mathbf{U}_v\subseteq\mathbf{V}$ of vertices that are, again, assigned to the same community in $\mathbf{x}^\ddagger$, $\mathbf{x}^\dagger$, or both and color the vertex $v$ in Figure \ref{six}e on the RGB color scale according to the fraction of vertices in $\mathbf{U}_v$ that fall into the analogous category in Figure \ref{six}a.  Vertices in the state of Indiana are highly red, indicating they were joined with a large community during the constraint-satisfaction procedure.  The large community with which the Indiana community merged included vertices from a sizable geographic region covering much of the Upper-Midwest region and, interestingly, a very small community consisting of twelve hospitals in the Southern-most region of Texas.

\addtocounter{section}{1}
\section*{\centering \arabic{section}. \uppercase{Extensions and Concluding Remarks}}
\addtocounter{section}{-1}
\refstepcounter{section}
\label{extrem}

Our method for identifying network communities which optimize a quality function while adhering to constraints establishes the utility of penalized optimization in community detection.  It is versatile and amenable to many types of constraints on the composition of communities.  We note that our procedure is valid for any constraint which is an increasing function of the variable of interest, e.g., ICD procedure volume, number of cardiac surgeons, quantity of cases involving improper medical procedures, etc.  The key requirement of our constrained optimization procedure is that the merging of two communities must not be the basis for the resulting community to be in violation of the constraint.  A constraint that imposes a maximum ICD volume is, for example, not of this type.

A useful extension that, in the context of modularity optimization, poses challenges is a constraint on the number of communities.  Due to the agglomerative approach of our procedure, there is no mechanism for reinserting communities that have gone extinct (Section \ref{locopt}).  We have experimented with an approach that, while not explicitly constraining the number of communities, maintains the same number of communities between the estimated unconstrained and constrained community assignments, e.g., $|\mathbf{x}^\ddagger|=|\mathbf{x}^\dagger|$.  We have approached this by identifying the communities in $\mathbf{x}^\ddagger$ which (i) violate the constraint and (ii) have excess ICD surgeries and so do not bind the constraint (Section \ref{model}).  Specifically, we match pairs of communities, one from each of (i) and (ii) above, and execute Algorithm \ref{alg3} sequentially on the subgraphs induced by each pair of communities. Exploration of this method is a topic of future research.

There exists a disconnect between network science and health services research due in part to the incongruence between mathematical elegance and real-world constraints. We have provided an illustration of the application of both a pure (unconstrained) method and one with constraints. We solved the practical problem of partitioning a network of hospitals with the constraint that the number of ICD surgical procedures that have taken place at hospitals belonging to each community exceeds some threshold. Though our method advances both the community detection and heath services literatures, it is not complete from the perspective of a health care policy maker since many real-world constraints remain to be incorporated.  Another type of constraint is, for example, given the geographic locations of hospitals, a requirement that communities not exceed a defined geographic maximum diameter or that they satisfy a geographic congruity constraint. This article is an initial exploration of a line of thinking that we anticipate will substantially advance the practical utility of
community detection.

In terms of health policy, our future research involving an outcomes-based analysis of the communities discovered by our method, as constrained here by minimum ICD surgery volume and subsequently by other factors, will lead to an improved understanding of characteristics of hospitals within communities and the detection of factors that drive variations in health care.  Through standardizing the composition of the HCCs, our method provides the tools for such comparisons to be made meaningfully.

\appendix
\section*{\centering \uppercase{Appendix: Enumerating Community Assignments}}\label{appendix}

The size of $\mathbf{K}_p=\{0,1,\ldots,p-1\}^p$ is clearly $|\mathbf{K}_p|=p^p$ which is much larger than necessary to contain all unique community assignments on a network of $p=|\mathbf{V}|$ vertices.  For example, if $\mathbf{x}\in\mathbf{K}_p$ then $(\mathbf{x}+1\mbox{ modulo } p)\in\mathbf{K}_p$ corresponds to the same community structure as $\mathbf{x}$.  Similarly, if $\mathbf{x}\in\mathbf{S}_\tau(\mathbf{V})$, for some $\tau\geq0$, then $(\mathbf{x}+1\mbox{ modulo } p)\in\mathbf{S}_\tau(\mathbf{V})$. Such multiplicity was inconsequential for our computational procedures and, in fact, maintaining a sample chain within the kernel of unique  community assignments would cause unnecessary checking and relabeling for diminished performance.  Never-the-less, to quantify the problem size and number of feasible solutions, we turn to well-known combinatorial numbers.  We utilize the \textit{Stirling numbers of the second kind} defined as
\begin{equation}
\stir{n}{k} = \frac{1}{k!}\sum_{j=0}^k(-1)^{k-j}\binom{k}{j}j^n
\end{equation}
which counts the number of ways to partition $n$ elements into $k$ groups.

Suppose $\tau=0$ so that the constraint that $F(v,\mathbf{x})>0$ for each $v\in\mathbf{V}$ is equivalent to the existence of at least one CCF per community.  Then, on a network of $p=|\mathbf{V}|$ vertices and $r=|\mathbf{V}^\prime|$ CCFs, there exists
\begin{equation}
|\mathbf{S}_0(\mathbf{V})| = \sum_{k=1}^r\stir{p-r}{k}\stir{r}{k}k!
\end{equation}
feasible community assignments.  Under the assumption $r<<p$, an upper bound $|\mathbf{S}_0(\mathbf{V})|$ is
\begin{equation}\label{sum}
|\mathbf{S}_0(\mathbf{V})| \leq \stir{p-r}{r}\sum_{k=1}^r\stir{r}{k}k!.
\end{equation}
The sum in Equation \eqref{sum} is known as the \textit{ordered Bell numbers} and is approximated as
\begin{equation}
\sum_{k=1}^r\stir{r}{k}k! \approx \frac{r!}{2(\log 2)^{r+1}}.
\end{equation}
Since, for a fixed value of $r$ the value $\stir{p-r}{r}\sim r^{p-r}/r!$, we approximate the number of feasible community assignments for $p\rightarrow\infty$ as
\begin{equation}\label{size}
|\mathbf{S}_0(\mathbf{V})| \lesssim \frac{r^{p-r}}{2(\log 2)^{r+1}}.
\end{equation}
The total number of isomorphically unique elements of $\mathbf{K}_p$ (which we denote as $\ker \mathbf{K}_p$), is given by
\begin{equation}
|\ker\mathbf{K}_p| = \sum_{k=1}^r\stir{p}{k} = \stir{p}{r}\sum_{k=1}^r\stir{p}{k}/\stir{p}{r} \sim \frac{r^p}{r!},
\end{equation}
again, for $p\rightarrow\infty$.  The fraction $|\mathbf{S}_0(\mathbf{V})|/|\ker\mathbf{K}_p|$ of the space $\ker\mathbf{K}_p$ occupied by $\mathbf{S}_0(\mathbf{V})$ is
\begin{equation}
\frac{|\mathbf{S}_0(\mathbf{V})|}{|\ker\mathbf{K}_p|} \lesssim \frac{ r!r^{-r}}{2(\log 2)^{r+1}} \approx \left(\frac{\sqrt{2\pi}}{2\log2}\right)\left(\frac{\sqrt{r}}{(e\log2)^r}\right),
\end{equation}
by applying Stirling's approximation to the factorial and constant factors to obtain $\sqrt{2\pi}/2\log2\approx 1.808$ and $e\log2\approx 1.884$, respectively.

This implies that, for large enough $r$, the constrained space occupies a diminishing fraction of the space of all community assignments.  While the reduction factor is significant, the size of the feasible space (Equation \ref{size}) remains exponential in the number of vertices in the network and computationally intractable to enumerate.

\bibliographystyle{ECA_jasa}

\end{document}